\documentclass[tightenlines,eqsecnum,floats,aps,amsmath,amssymb,nofootinbib,prd,showpacs]{revtex4}
\pdfoutput=1 
\usepackage{amsmath,amssymb,amsfonts}
\usepackage{graphicx}
\usepackage{enumerate} 
\usepackage{colordvi} 
\usepackage{bm}
\usepackage{epsfig}
\usepackage{float}
\usepackage{verbatim}
\usepackage{subfig}
\usepackage{natbib}

\newcommand{\bfv}{\mbox{\boldmath$v$}}

\newcommand{\bfk}{\mbox{\boldmath$k$}}

\begin{document}
\title{A Perturbative Approach to the Redshift Space Power Spectrum: Beyond the Standard Model}

\vfill
\author{Benjamin Bose$^{1}$, Kazuya Koyama$^{1}$}
\bigskip
\affiliation{$^1$Institute of Cosmology \& Gravitation, University of Portsmouth,
Portsmouth, Hampshire, PO1 3FX, UK}
\bigskip
\vfill
\date{today}

\begin{abstract}
We develop a code to produce the power spectrum in redshift space based on standard perturbation theory (SPT) at 1-loop order. The code can be applied to a wide range of modified gravity and dark energy models using a recently proposed numerical method by A.Taruya to find the SPT kernels. This includes Horndeski's theory with a general potential, which accommodates both chameleon and Vainshtein screening mechanisms and provides a non-linear extension of the effective theory of dark energy up to the third order.  Focus is on a recent non-linear model of the redshift space power spectrum which has been shown to  model the anisotropy very well at  relevant scales for the SPT framework, as well as capturing relevant non-linear effects typical of modified gravity theories. We provide consistency checks of the code against established results and elucidate its application within the light of upcoming high precision RSD data.   
\end{abstract}
\maketitle

\section{Introduction}
One of the most recent and most perplexing problems in physics  is that of the late time acceleration of the Universe discovered at the turn of the last century \cite{Riess:1998cb,Perlmutter:1998np}. If we take Einstein's theory of general relativity (GR) to be the correct theory of gravity, this acceleration can be accounted for through a cosmological constant whose physical interpretation is that of a dark energy component. The major problems with this picture are the fact that this energy is directly unobserved and moreover, the amount of acceleration we observe implies an exceedingly small value for this constant leading to the famous fine tuning problem, e.g. \cite{Weinberg:1988cp,Martin:2012bt}. One may also consider the so called 'Coincidence Problem', which asks why the dark energy and matter components of this long lived, dynamical universe  are comparable today .These problems naturally led physicists to consider alternative solutions to the observations. Modifying GR is one of the most natural paths to that end and has been investigated thoroughly, e.g., \cite{Hu:2007nk,Starobinsky:1980te,Dvali:2000hr,Gomes:2013ema,Hassan:2011zd,deRham:2011by,deRham:2010ik,deRham:2010kj,Maartens:2010ar}. 
\newline
\newline
The main problem facing these attempts is the local success of GR. In general, by modifying GR we introduce an extra force coming from the extra degree of freedom. This so called fifth force is not observed in any solar system tests. This can be solved by introducing a screening mechanism which allows any modifications to be filtered out at solar system scales (See \cite{Koyama:2015vza,Sakstein:2015oqa,Clifton:2011jh} for reviews). These mechanisms come in two general forms, screening the force's source or screening through non-linearities. The former comes as either the symmetron mechanism \cite{Hinterbichler:2010es} or the Chameleon mechanism \cite{Khoury:2003aq,Khoury:2003rn,Hu:2007nk,Brax:2008hh}, while the latter is called Vainshtein screening \cite{Vainshtein:1972sx}. All these mechanisms can be accommodated in Horndeski's most general scalar-tensor theory \cite{Horndeski:1974wa} with a generalised potential term. This is the level of generality we aim at in this work. 
\newline
\newline
Having required deviations from GR be undetectable locally through screening, in order to test these modifications to GR we must get out of the solar system where screening is weak. This makes the large scale structure of the universe an ideal laboratory for probing these models of gravity and much work has been done to this end \cite{Uzan:2000mz,Lue:2004rj,Ishak:2005zs,Knox:2005rg,Koyama:2006ef,Chiba:2007rb,Amendola:2007rr,Simpson:2012ra,Terukina:2012ji,Terukina:2013eqa,Yamamoto:2010ie,Jain:2007yk,Zhao:2008bn,Zhao:2009fn,Asaba:2013xql}. 
\newline
\newline
In particular we consider the systematic effect known as redshift space distortions (RSD) \cite{Kaiser:1987qv}. This effect relates the peculiar velocities of galaxies to an observed anisotropy in redshift space.  Fifth force effects  directly relate to the peculiar velocities of galaxies making RSD a probe of great potential and with the ever increasing precision of galaxy surveys it becomes ever more promising in its ability to distinguish between models ,e.g. \cite{Linder:2007nu,Guzzo:2008ac,Yamamoto:2008gr,Song:2008qt,Song:2010bk,Guzik:2009cm,Song:2010fg,Asaba:2013mxj}. It is important to note that apart from redshift space information one also requires geometrical information to have a proper map between angle/redshift  and physical distance. This information is available from standard ruler or candle measurements such as the Baryon Oscillation Spectroscopic Survey within SDSS III \footnote{http://www.sdss3.org}, eBOSS projects \footnote{www.sdss.org/surveys/eboss/} and DESI  \footnote{\url{http://desi.lbl.gov/}}.
\newline
\newline 
Consequently, with the heightened precision of recent and upcoming surveys, theory is required to keep up in terms of its predictive power. Even though made at linear to quasi linear scales, non-linearities of the gravitational interaction play an important role in the RSD observations  and a lot of work has been done on modelling RSD beyond the linear Kaiser model, e.g.\cite{Matsubara:2007wj,Scoccimarro:2004tg,Heavens:1998es,Percival:2008sh,Magira:1999bn,Taruya:2010mx}. This non-linear dependence makes it crucial to properly model RSD when changing the structure of gravity in order to get the correct and distinguishing predictions. In this work we give attention to an improved RSD model developed in Ref.\cite{Taruya:2010mx} which we call the TNS model after its authors.  The model uses standard perturbation theory (SPT) to construct the non-linear redshift space power spectrum and has been shown to perform very well when compared to N-body simulations \cite{Nishimichi:2011jm,Taruya:2013my,Ishikawa:2013aea}. 
\newline
\newline
It is worth discussing the overlap and distinction between modified gravity effects and the flexibility of standard GR templates, i.e. nuisance parameters such as galaxy or halo bias, when we consider the redshift space power spectrum. Whether the standard GR treatment gives biased results for the constraints on modified gravity model parameters or not depends on the method used in modelling RSD and also the precision of the measurements. Clearly, if the errors in the measurements are larger than the bias introduced by the inaccurate modelling of RSD, there is no need to improve the RSD modelling. However this needs to be tested before the method is applied to real data. For the TNS model, which has been used to measure $f \sigma_8$ using the power spectrum measurements of BOSS survey \cite{Beutler:2016arn}, this has been done only for $f(R)$ gravity. In this model, it was shown that the standard GR template gives a biased estimation of the model parameter assuming an ideal survey with volume $10 Gpc^3 h^{-3} $ at $z=1$  \cite{Taruya:2013my}. Thus, it is important to test whether this kind of bias exists for other MG models. In Ref.\cite{Barreira:2016mg} they find that with their analysis of the redshift space power spectrum for a Vainshtein screened model, no model bias is found. The expectation is that this absence of model bias is true for scale independent models of gravity (unlike the scale dependent $f(R)$ case). In a future work we will test for consistency with this result. For this, clearly we need to develop the TNS model including the modified gravity effect consistently. 
\newline
\newline
Recently a method to find the SPT kernels numerically was developed in Ref.\cite{Taruya:paper}. In the work the authors apply this method to calculate the real space power spectrum for GR and the $f(R)$ model of gravity. Here, we present a new code which follows their method for computing the perturbative kernels up to 3rd order and then extends their work in applying the kernels to calculate the TNS redshift space power spectrum. The code also uses the general applicability of the numerical method to accommodate a large class of gravity and dark energy models. In particular, we describe and show results for our code's application to a Vainshtein screened and chameleon screened model, namely the DGP model of gravity \cite{Dvali:2000hr} and the Hu-Sawicki form of $f(R)$ gravity \cite{Hu:2007nk}. 
\newline
\newline
This paper is organised as follows: In Sec.II we review the relevant theory. We start by deriving the generalised evolution equations for the velocity and density perturbations, highlighting where the modifications to GR arise, parametrizing these modifications  in terms of 1st, 2nd and 3rd order effects. We also present expressions for the 1-loop auto and cross real space power spectra in terms of the 1st , 2nd and 3rd order perturbation kernels. A Chameleon and Vainshtein screened model are given as explicit examples under our parametrisation. We finish the section with a review of the TNS model of RSD, giving the analytical expressions in terms of our generalised SPT kernels. In Sec.III we detail our code and provide some results. First we compare the code's results for the 1-loop real space power spectrum and TNS monopole with well established analytical results in SPT for both GR and a Vainshtein screened model of gravity. This is done as a consistency check. Next we compare the code's results with N-body results for a Chameleon screened model of gravity in an attempt to reproduce some of the results of Ref.\cite{Taruya:2013quf}. Finally, we summarise our results and highlight future work in Sec.V.

\section{Theory}
\subsection{Evolution equations for the perturbations}
We consider perturbations around the Friedman-Robertson-Walker (FRW)
universe described in the Newtonian gauge.
\begin{equation}
ds^2=-(1+2\Phi)dt^2+a^2(1-2\Psi)\delta_{ij}dx^idx^j.
\end{equation}
We will work on the evolution of matter fluctuations inside the Hubble horizon.
Then we can use the quasi-static approximation and neglect the time derivatives
of the perturbed quantities compared with the spatial derivatives. As mentioned
in the introduction, the large distance modification of gravity which is
necessary to explain the late-time acceleration generally modifies gravity
even on sub-horizon scales due to the introduction of a new scalar degree
of freedom. In this paper, we use the Jordan frame where matter is minimally coupled to gravity. The evolution equations for matter perturbations are obtained from the conservation of the energy momentum tensor. This gives the continuity
and Euler equations.
\begin{eqnarray}
&&\frac{\partial \delta}{\partial t}+\frac{1}{a}\nabla\cdot
[(1+\delta)\bfv]=0,
\label{eq:continuity_eq}\\
&&\frac{\partial \bfv}{\partial t}+H\bfv+\frac{1}{a}(\bfv\cdot\nabla)\bfv=
-\frac{1}{a}\nabla \Phi.
\label{eq:Euler_eq}
\end{eqnarray}

Assuming the irrotationality of fluid quantities, the velocity field
$\bfv$ is expressed in terms of velocity divergence
$\theta\equiv\nabla\cdot\bfv/(a\,H)$. Then the Fourier transform of the
fluid eq.(\ref{eq:continuity_eq}) and eq.(\ref{eq:Euler_eq}) become
\begin{eqnarray}
&&a \frac{\partial \delta(\bfk)}{\partial a}+\theta(\bfk) =-
\int\frac{d^3\bfk_1d^3\bfk_2}{(2\pi)^3}\delta_{\rm D}(\bfk-\bfk_1-\bfk_2)
\alpha(\bfk_1,\bfk_2)\,\theta(\bfk_1)\delta(\bfk_2),
\label{eq:Perturb1}\\
&& a \frac{\partial \theta(\bfk)}{\partial a}+
\left(2+\frac{a H'}{H}\right)\theta(\bfk)
-\left(\frac{k}{a\,H}\right)^2\,\Phi(\bfk)=
-\frac{1}{2}\int\frac{d^3\bfk_1d^3\bfk_2}{(2\pi)^3}
\delta_{\rm D}(\bfk-\bfk_1-\bfk_2)
\beta(\bfk_1,\bfk_2)\,\theta(\bfk_1)\theta(\bfk_2),
\label{eq:Perturb2}
\end{eqnarray}
the prime denoting a scale factor derivative and the kernels in the Fourier integrals, $\alpha$  and
$\beta$, are given by
\begin{eqnarray}
\alpha(\bfk_1,\bfk_2)=1+\frac{\bfk_1\cdot\bfk_2}{|\bfk_1|^2},
\quad\quad
\beta(\bfk_1,\bfk_2)=
\frac{(\bfk_1\cdot\bfk_2)\left|\bfk_1+\bfk_2\right|^2}{|\bfk_1|^2|\bfk_2|^2}.
\label{alphabeta}
\end{eqnarray}
The information on modifications of gravity is encoded in the relation between the Newtonian potential $\Phi$ and $\delta$. In order to calculate the non-linear power spectrum at 1-loop order using SPT, we need to expand $\delta$ up to the third order using SPT (See Ref.\cite{Bernardeau:2001qr} for a review). We assume the Newtonian potential is related to $\delta$ non-linearly and we expand this relation up to the third order \cite{Koyama:2009me}
\begin{equation}
-\left(\frac{k}{a H}\right)^2\Phi=
\frac{3 \Omega_m(a)}{2} \mu(k,a)\,\delta(\bfk) + S(\bfk),
\label{eq:poisson1}
\end{equation}
where $\Omega_m(a) = 8 \pi G \rho_m/3 H^2$.
The function $S(\bfk)$ is the non-linear source term up to the third order given by
\begin{eqnarray}
S(\bfk)&=&
\int\frac{d^3\bfk_1d^3\bfk_2}{(2\pi)^3}\,
\delta_{\rm D}(\bfk-\bfk_{12}) \gamma_2(\bfk, \bfk_1, \bfk_2;a)
\delta(\bfk_1)\,\delta(\bfk_2)
\nonumber\\
&& + 
\int\frac{d^3\bfk_1d^3\bfk_2d^3\bfk_3}{(2\pi)^6}
\delta_{\rm D}(\bfk-\bfk_{123})
\gamma_3( \bfk, \bfk_1, \bfk_2, \bfk_3;a)
\delta(\bfk_1)\,\delta(\bfk_2)\,\delta(\bfk_3)
\label{eq:Perturb3}
\end{eqnarray}
where $\gamma_2( \bfk, \bfk_1, \bfk_2; a)$  and $\gamma_3(\bfk, \bfk_1, \bfk_2, \bfk_3;a)$ are symmetric under the exchange of $\bfk_i$. 
\newline
\newline
We solve eq.(\ref{eq:Perturb1}) and eq.(\ref{eq:Perturb2}) perturbatively. The n-th order solutions are given by 
\begin{align} 
\delta_n(\boldsymbol{k}, a) &= \int d^3\boldsymbol{k}_1...d^3 \boldsymbol{k}_n \delta_D(\boldsymbol{k}-\boldsymbol{k}_{1...n}) F_n(\boldsymbol{k}_1,...,\boldsymbol{k}_n, a) \delta_0(\boldsymbol{k}_1)...\delta_0(\boldsymbol{k}_n) \label{nth1} \\ 
\theta_n(\boldsymbol{k},a) &= \int d^3\boldsymbol{k}_1...d^3 \boldsymbol{k}_n \delta_D(\boldsymbol{k}-\boldsymbol{k}_{1...n}) G_n(\boldsymbol{k}_1,...,\boldsymbol{k}_n, a) \delta_0(\boldsymbol{k}_1)...\delta_0(\boldsymbol{k}_2) \label{nth2}
\end{align}
where $\boldsymbol{k}_{1...n} = \boldsymbol{k}_1 + ...+ \boldsymbol{k}_n$. In the following we will omit the scale factor dependence of the kernels to simplify the expressions. 
\newline
\newline
By insterting eq.(\ref{nth1}) and eq.(\ref{nth2}) into the continuity and Euler equations we get a generalised form of the system for the $n^{th}$ order kernel  (See Ref. \cite{Taruya:paper} for more details)
\begin{align}
  \hat{\mathcal{L}}(k)  \left( \begin{array}{c}
\emph{F}_n(\boldsymbol{k_1}, \cdot \cdot \cdot , \boldsymbol{k_n}) \\
\emph{G}_n(\boldsymbol{k_1}, \cdot \cdot \cdot , \boldsymbol{k_n}) \end{array} \right) = 
\sum^{n-1}_{j=1}
\left( \begin{array}{c}
- \emph{$\alpha$}(\boldsymbol{k_{1\cdot \cdot \cdot j}},  \boldsymbol{k_{j+1 \cdot \cdot \cdot n}} ) 
G_j(\boldsymbol{k_1},\cdot \cdot \cdot, \boldsymbol{k_j}) F_{n-j}(\boldsymbol{k_{j+1}},\cdot \cdot \cdot ,\boldsymbol{k_n})  \\
-\frac{1}{2}\emph{$\beta$}(\boldsymbol{k_{1\cdot \cdot \cdot j}}, \boldsymbol{k_{j+1 \cdot \cdot \cdot n}}) G_j(\boldsymbol{k_1}\cdot \cdot \cdot \boldsymbol{k_j}) G_{n-j}(\boldsymbol{k_{j+1} }, \cdot \cdot \cdot, \boldsymbol{k_n}) - N_n(\bfk, \boldsymbol{k_1}, \cdot \cdot \cdot, \boldsymbol{k_n})   
 \end{array} \right)  
 \label{eq:dynamicsystem}
\end{align}
where 
\begin{equation}
  \hat{\mathcal{L}}(k) = \left( \begin{array}{cc}
a\frac{d}{da} & 1 \\
\frac{3 \Omega_m}{2} \mu(k,a)   & a\frac{d}{da}+\left(2+\frac{aH'}{H}\right) 
 \end{array} \right) 
\end{equation}
and 
\begin{align}
N_2 &= \emph{$\gamma$}_2(\bfk, \boldsymbol{k_{1}}, \boldsymbol{k_{2}}) F_1(\boldsymbol{k_1}) F_{1}(\boldsymbol{k_{2}}) \\
N_3 &= 
\emph{$\gamma$}_2(\bfk, \boldsymbol{k_{1}},\boldsymbol{k_{23}})
F_1(\boldsymbol{k_{1}}) F_2(\boldsymbol{k_{2}}, \boldsymbol{k_{3}})
+\emph{$\gamma$}_2(\bfk, \boldsymbol{k_{12}},\boldsymbol{k_{3}})
F_2(\boldsymbol{k_{1}}, \boldsymbol{k_{2}}) F_1(\boldsymbol{k_{3}})
+ \emph{$\gamma$}_3(\bfk, \boldsymbol{k_{1}}, \boldsymbol{k_{2}}, \boldsymbol{k_{3}})
 F_1(\boldsymbol{k_1}) F_{1}(\boldsymbol{k_{2}}) F_{1}(\boldsymbol{k_{3}}).
\end{align}
Since $\bfk_1, ...,\bfk_n$ are integration variables, we can symmetrise $F_n$ and $G_n$ with respect to the exchange of $\bfk_1, ..., \bfk_n$. In the following we symmetrise the kernels by symmetrising $\alpha(\bfk_1, \bfk_2)$ and adding cyclic permutations to the right hand side.  From eq.(\ref{eq:dynamicsystem}) we note that the 1st order kernels can at most spatially depend on $k$ so we will write $F_1(\bfk_i) = F_1(k)$ and $G_1(\bfk_i)=G_1(k)$. 
\newline 
\newline
The one-loop power spectra are given by
\begin{align}
\langle g^i_{2}(\bfk) g^{j}_2(\bfk')\rangle &=
(2\pi)^3\delta_{\rm D}(\bfk+\bfk')\,P^{ij}_{22}(k). \label{eq:psconstraint0} \\
\langle g^i_{1}(\bfk) g^{j}_3(\bfk')
+g^i_{3}(\bfk) g^{i}_1(\bfk') \rangle &=
(2\pi)^3\delta_{\rm D}(\bfk+\bfk')\,P^{ij}_{13}(k).
\label{eq:psconstraint1}
\end{align}
where $g^1 = \delta$ and $g^2= \theta$. In terms of the 2nd and 3rd order kernels they are given as 
\begin{align}
P^{\delta \delta}_{22}(k)
&= 2 \int d^3 \bfk' F_2 (\bfk - \bfk', \bfk')^2 
P_0(|\bfk - \bfk'|) P_0(k'), \\
P^{\delta \theta}_{22}(k) 
&= 2 \int d^3 \bfk' F_2 (\bfk - \bfk', \bfk') G_2 (\bfk - \bfk', \bfk')
P_0(|\bfk - \bfk'|) P_0(k'), \\
P^{\theta \theta}_{22}(k) 
& = 2 \int d^3 \bfk' G_2 (\bfk - \bfk', \bfk')^2
P_0(|\bfk - \bfk'|) P_0(k'),
\end{align}
and 
\begin{align}
P^{\delta \delta}_{13}(k) 
&= 6 \int d^3 \bfk' F_3(\bfk, \bfk', -\bfk') 
P_0(k')  F_1(k) P_0(k), \\
P^{\delta \theta}_{13}(k) 
&= 3  \int d^3 \bfk' G_3 (\bfk, \bfk', -\bfk')
P_0(k')  F_1(k) P_0(k) + 3  \int d^3 \bfk' F_3 (\bfk, \bfk', -\bfk') 
P_0(k') G_1(k) P_0(k), \\
P^{\theta \theta}_{13}(k) 
& = 6 \int d^3 \bfk' G_3 (\bfk, \bfk', -\bfk') 
P_0(|\bfk - \bfk'|) G_1(k) P_0(k'),
\end{align}
where $P_0(k)$ is defined as 
\begin{equation}
\langle \delta_0(\bfk) \delta_0(\bfk')\rangle =
(2\pi)^3\delta_{\rm D}(\bfk+\bfk')\,P_0(k).
\end{equation}
By defining $k'= k r$ and $\bfk \cdot \bfk' = k^2 r x$, these integrations can be written as 

\begin{align}
P^{\delta \delta}_{22}(k) & = 2\frac{k^3}{(2\pi)^2} \int_0^{\infty}  dr r^2 \int^{1}_{-1} dx P_0(kr) P_0(k\sqrt{1+r^2-2rx}) F_2(k,r,x)^2 \\ 
P^{\delta \delta}_{13}(k) & = 2\frac{k^3}{(2\pi)^2} F_1(k) P_0(k)  \int_{0}^{\infty}  dr r^2 P_0(kr)  F_3(k,r,x)
\label{eq:loopkernels}
\end{align}
where we defined $F_2(\bfk-\bfk', \bfk') = F_2(k,r, x)$, $F_3(\bfk, \bfk', -\bfk')=F_3(k,r, x)$. 

\subsection{Examples}
Here we provide some examples using the above parametrisation. To start, for GR we simply have $\mu(k;a) =1$ and $\gamma_2(\bfk, \bfk_1,\bfk_2 ;a) = \gamma_3(\bfk, \bfk_1, \bfk_2, \bfk_3;a) = 0$. Next we provide an example of a Chameleon and a Vainshtein screened model. In the Appendix we provide the parametrisation for Horndeski's most general scalar tensor theory with a generalised potential term. 

\subsubsection{$f(R)$} 
$f(R)$ gravity (See \cite{Sotiriou:2008rp,Capozziello:2007ec} for reviews) is one of the most straightforward extensions to Einstein's theory, replacing the scalar curvature $R$ in the Einstein Hilbert action with a general function $f(R)$

\begin{equation}
S = \int d^4x \sqrt{-g} \left[ \frac{R +f(R)}{2\kappa^2} + L_m \right]
\end{equation}
where $\kappa^2 = 8\pi G$ and $L_m$ is the matter Lagrangian. The theory is known to be equivalent to a scalar tensor theory with a non trivial potential, which becomes evident when we look at the trace of the modified Einstein field equations
\begin{equation}
3\Box f_R - R +f_R R -2f(R) = -\kappa^2 \rho_m
\end{equation}
where $f_R = df(R)/dR$ and $\rho_m$ is the matter density of the universe (we have assumed a purely matter dominated universe). Specifically, we give the example of the Chameleon screened Hu-Sawicki form of $f(R)$ \cite{Hu:2007nk}, given by 
\begin{equation}
f(R)  = -m^2 \frac{c_1 (R/m^2)^n}{c_2(R/m^2)^n+1}
\end{equation}
and when we set $n=1$ we have 
\begin{equation}
f(R)  \propto \frac{R}{AR+1}
\end{equation} 
with $A$ being a constant with dimensions of length squared. In a high curvature regime $AR >> 1$ we can expand $f(R)$ as 
\begin{equation}
f(R) \simeq -2\kappa^2 \rho_\Lambda - f_{R0} \frac{R_0^2}{R}
\end{equation}
where $\rho_\Lambda$ depends on $A$, $R_0$ is the background curvature today and $f_{R0}=\bar{f}_R(R_0)$, the bar indicating it is evaluated on the background. $|f_{R0}|$ is the free parameter of the theory. If we want to follow a LCDM background we find that 
\begin{equation}
R_0 = H_0^2(12-9\Omega_{m0})
\end{equation}
$H_0$ and $\Omega_{m0}$ being the Hubble parameter and matter density fraction today. Using the above relations and the $f(R)$ form of the Poisson equation (See Ref.\cite{Koyama:2009me} for example), we can compare with Eq.\ref{eq:poisson1} to get following non-linear interaction terms  

\begin{equation}
\mu(k;a) = 1 + \left(\frac{k}{a}\right)^2\frac{1}{3\Pi(k;a)}
\end{equation}

\begin{equation}
\gamma_2(k,\bfk_1,\bfk_2;a)  =- \frac{9}{48}\left(\frac{kH_0}{aH}\right)^2\left(\frac{H_0^2\Omega_{m0}}{a^3}\right)^2\frac{(\Omega_{m0} -4a^3(\Omega_{m0}-1))^5}{a^{15}|f_{R0}|^2 (3\Omega_{m0}-4)^4}\frac{1}{\Pi(k;a)\Pi(k_1;a)\Pi(k_2;a)} 
\end{equation}
\begin{align}
\gamma_3(k,\bfk_1,\bfk_2,\bfk_3;a)  & =  \left(\frac{kH_0}{aH}\right)^2 \left(\frac{H_0^2\Omega_{m0}}{a^3}\right)^3 \frac{1}{36\Pi(k;a)\Pi(k_1;a)\Pi(k_2;a)\Pi(k_3;a)\Pi(k_{23};a)}  \nonumber \\ 
& \times  \left[-\frac{45}{8} \frac{\Pi(k_{23};a)}{a^{21}|f_{R0}|^3}\left( \frac{(\Omega_{m0} - 4a^3(\Omega_{m0}-1))^7}{(3\Omega_{m0}-4)^6} \right)+H_0^2\left( \frac{9}{4a^{15} |f_{R0}|^2} \frac{(\Omega_{m0}-4a^3(\Omega_{m0}-1))^5}{(3\Omega_{m0}-4)^4} \right)^2\right]
\end{align}
where
\begin{equation}
\Pi(k;a) = \left(\frac{k}{a}\right)^2+\frac{H_0^2(\Omega_{m0} - 4a^3(\Omega_{m0}-1))^3}{2|f_{R0}|a^9(3\Omega_{m0}-4)^2}
\end{equation}

\subsubsection{DGP}
\noindent In the DGP model of gravity \cite{Dvali:2000hr}  we live on a 4 dimensional brane embedded in 5 dimensional Minkowski spacetime, giving this theory a crossover scale $r_c$, which is the ratio between the 5D Newton gravitational constant and the 4D Newton  gravitational constan. $r_c$ is the free parameter of the theory and we parametrize it as $\Omega_{rc}= 1/(4r_c^2H_0^2)$, $H_0$ being the Hubble parameter today. The modified Friedman equation in this theory is given by 
\begin{equation}
\epsilon\frac{H}{r_c} = H^2(1 - \Omega_m(a))
\end{equation}
where $\epsilon=\pm 1$. The solution for $\epsilon=1$ is known to be ghostly and so we consider what is called the normal branch with $\epsilon = -1$. In this branch acceleration is achieved through a dark energy constant as in GR. We also impose a background history following LCDM, done by tuning the dark energy equation of state. The non-linear interaction terms for this theory are found to be (for details on the Poisson equation form in this theory see Ref.\cite{Koyama:2009me} for example)

\begin{equation}
\mu(k;a)  = 1 + \frac{1}{3\beta(a)} 
\end{equation}
\begin{equation}
\gamma_2(k,\bfk_1,\bfk_2;a) = -\frac{H_0^2}{24 H^2 \beta(a)^3 \Omega_{rc}} \left(\frac{\Omega_{m0}}{a^3}\right)^2 (1-\mu_{1,2}^2) 
\end{equation}
\begin{equation}
\gamma_3(k,\bfk_1,\bfk_2,\bfk_3;a) = \frac{H_0^2}{144 H^2 \beta(a)^5 \Omega_{rc}^2} \left(\frac{\Omega_{m0}}{a^3}\right)^3 (1-\mu_{1,2}^2) (1-\mu_{1,23}^2)
\end{equation}
where
\begin{equation}
\beta(a)= 1+\frac{H}{\Omega_{rc}}\left(1+\frac{aH'}{3H}\right)
\end{equation}
$\mu_{i,j}$ is the cosine of the angle between $\bfk_i$ and $\bfk_j$, $\bfk_{ij}=\bfk_i+\bfk_j$ and $H'=dH/da$.  

\subsubsection{Horndeski's Theory}
In this section we have provided two specific models as examples. Both of these can be included or compared to Horndeski's general framework with a general scalar field potential. This class of theories is the most general class of theories including an extra scalar degree of freedom and whose equations of motion are at most 2nd order, and the theories under its umbrella are free from  the Ostrogradsky instability (See Ref.\cite{Woodard:2006nt} for a discussion). In Appendix. B we provide  this general framework and relate it to the modified Euler equation parameters $\mu(k;a)$, $\gamma_2(\bfk, \bfk_1,\bfk_2;a)$ and $\gamma_3(\bfk, \bfk_1,\bfk_2,\bfk_3;a)$. 


\subsection{Redshift Space Distortions}
As mentioned in the introduction, RSD arises from the non-linear mapping between real and redshift space due to the particular velocities of galaxies. At the linear level we have a squashing of the galaxy distribution coming from motion towards the distribution's centre. This was first modelled by Kaiser \cite{Kaiser:1987qv} 
\begin{equation}
P_{K}^S(k,\mu) = (1+\mu^2)^2P_{\delta \delta}(k)
\label{linkais}
\end{equation}
where $\mu$ \footnote{The use of $\mu$ here should not be confused with the function $\mu(k;a)$ which will always include its arguments.} is the cosine of the angle between the line of sight and $\bfk$ and $P_{\delta \delta}(k)$ is the linear matter power spectrum. A non-linear version of the Kaiser model was later proposed introducing the interactions between velocities and density via the velocity auto and cross power spectra, $P_{\theta \theta}$ and $P_{\delta \theta}$ respectively. 
\begin{equation}
P_{K}^S(k,\mu) = P_{\delta \delta}(k) +2\mu^2  P_{\delta \delta}(k)  + \mu^4 P_{\delta \delta}(k) 
\label{nonlinkais}
\end{equation}
Both models still did not account for the small scale damping effect of random virialised motion, or Fingers of God (FoG) effect. Many authors accounted this effect via a phenomenological pre factor term, usually taking the form of an exponential or Gaussian \cite{Scoccimarro:2004tg,Percival:2008sh,Cole:1994wf,Peacock:1993xg,Park:1994fa,Ballinger:1996cd,Magira:1999bn}. 
\newline
\newline
Recently, an improved non-linear model of RSD was  presented in Ref.\cite{Taruya:2010mx}. From conservation of energy one can derive an exact general expression for the redshift-space power spectrum
\begin{equation}
P^S(\boldsymbol{k}) = \int d^3 \boldsymbol{x} e^{i \boldsymbol{k} \cdot \boldsymbol{x}} \langle e^{j_1 A_1} A_2 A_3 \rangle
\label{PSred1}
\end{equation}
with 
\begin{align}
j_1= &-i k \mu \\ \nonumber 
A_1 = &u_z(\boldsymbol{r}) - u_z(\boldsymbol{r}')  \\ \nonumber 
A_2 = &\delta(\boldsymbol{r})  -\nabla_z u_z(\boldsymbol{r})  \\ \nonumber 
A_3 = &\delta(\boldsymbol{r}') - \nabla_z u_z(\boldsymbol{r}') \nonumber 
\end{align}
where $\boldsymbol{x}=\boldsymbol{r}-\boldsymbol{r}'$ is the separation in real space and $\mu = (\bfk \cdot \boldsymbol{\hat{z}})/k$ is the directional cosine between the wave vector $\boldsymbol{k}$ and the $z$-axis which is taken to be the line of sight axis.  We also define  $u_z(\boldsymbol{r})=  v_z(\boldsymbol{r})/(a H)$  as the line of sight component of the velocity field. The only approximation made in deriving the above expression is the distant-observer-approximation which is valid for any survey that subtends a small angle on the sky. This is usually a good approximation for the surveys and redshifts we are interested in. 
\newline
\newline
The next step is to write this expression in terms of cumulants (See Ref.\cite{Taruya:2010mx}). Upon doing this, an overall exponential factor of the form exp$\{\langle e^{ik\mu A_1}\rangle_c \}$, with $\langle \cdots \rangle_c$ denoting the cumulant, is found. This factor is known to dampen the redshift space power spectrum at non-linear scales due to the virialized, highly uncorrelated motion. The character of this factor is known to be partly non-perturbative but it has been shown to mainly change the overall shape of the power spectrum and so it can be replaced  with the phenomenological functional form of $\mbox{D}_{\mbox{FoG}}(k\mu f \sigma_v)$ with the fitting parameter $\sigma_v$ which represents the velocity dispersion of the galaxy distribution. We will refer to this function as the FoG function. 
\newline
\newline
The quantities $A_2$ and $A_3$ are the terms leading to the Kaiser effect \cite{Kaiser:1987qv,Hamilton:1992zz}. The cumulant terms including these also include a factor of $e^{j_1 A_1}$  whose contribution should be small on the scales we consider. Given this,  we can  apply a perturbative treatment  to the expression. Stopping at quadratic order in the linear power spectrum, we get the following expression for the anisotropic redshift space power spectrum  
 \begin{equation}
 P^S(k,\mu) = \mbox{D}_{\mbox{FoG}} (k\mu f \sigma_v) \{ P_{\delta \delta} (k) - 2  \mu^2 P_{\delta \theta}(k) + \mu^4 P_{\theta \theta} (k) + A(k,\mu) + B(k,\mu) \} 
 \label{redshiftps}
 \end{equation}
 \noindent where $P_{\delta \delta}, P_{\delta \theta}$ and $ P_{\theta \theta}$ are all at 1-loop order. Also, $f = d \ln{F_1(k;a)}/d \ln{a}$ is the logarithmic growth factor. The correction terms, A and B are given by 
 \begin{equation}
 A(k,\mu)=  -(k \mu) \int d^3 \boldsymbol{k'} \frac{k_z '}{k'^2} \{B_\sigma(\boldsymbol{k'},\boldsymbol{k}-\boldsymbol{k'},-\boldsymbol{k})-B_\sigma(\boldsymbol{k'},\boldsymbol{k}, -\boldsymbol{k}-\boldsymbol{k'}) \} 
 \label{Aterm}
 \end{equation}
 \begin{equation}
 B(k,\mu)= (k \mu)^2 \int d^3\boldsymbol{k'} F(\boldsymbol{k'}) F(\boldsymbol{k}-\boldsymbol{k'})
 \label{Bterm}
 \end{equation}
 where
 \begin{equation}
 F(\boldsymbol{k}) = \frac{k_z}{k^2}\left[P_{\delta \theta} (k) - \frac{k_z^2}{k^2}P_{\theta \theta} (k) \right] 
 \end{equation}
 The cross bispectrum $B_\sigma$ is given by
 \begin{equation}
 \delta_D(\boldsymbol{k}_1+ \boldsymbol{k}_2+ \boldsymbol{k}_3)B_\sigma( \boldsymbol{k}_1,\boldsymbol{k}_2,\boldsymbol{k}_3) = \langle \theta(\boldsymbol{k}_1)\{ \delta(\boldsymbol{k}_2) - \frac{k_{2z}^2}{k_2^2} \theta(\boldsymbol{k}_2)\}\{ \delta(\boldsymbol{k}_3) - \frac{k_{3z}^2}{k_3^2} \theta(\boldsymbol{k}_3)\}
 \label{lcdmbi}
 \end{equation}
\newline
We can write  $B_\sigma$  up to 2nd order in the linear power spectrum by expanding the perturbations up to 2nd order. What we obtain is the following

\begin{align}
 B_\sigma(\boldsymbol{k}_1,\boldsymbol{k}_2,\boldsymbol{k}_3) &= 2\left[ \left(F_1(k_2) -\frac{k_{2z}^2}{k_2^2} G_1(k_2) \right)\left(F_1(k_3) -\frac{k_{3z}^2}{k_3^2} G_1(k_3)\right) G_2(\bfk_2,\bfk_3) P_0(k_2) P_0(k_3) \right. \nonumber \\ 
 &G_1(k_1)\left(F_1(k_3) -\frac{k_{3z}^2}{k_3^2} G_1(k_3)\right) \left(F_2(\bfk_1, \bfk_3) -\frac{k_{2z}^2}{k_2^2} G_2(\bfk_1,\bfk_3)\right)P_0(k_1) P_0(k_3) \nonumber \\ 
 & \left. G_1(k_1)\left(F_1(k_2) -\frac{k_{2z}^2}{k_2^2} G_1(k_2)\right) \left(F_2(\bfk_1, \bfk_2) -\frac{k_{3z}^2}{k_3^2} G_2(\bfk_1,\bfk_2)\right)P_0(k_1) P_0(k_2)  \right]
 \label{bispectrum}
 \end{align}

\noindent where $P_0(k)$ is the very early time linear power spectrum. It is useful to note here that using  $B_\sigma$'s symmetries, $B_\sigma(\bfk_1,\bfk_2,\bfk_3) = B_\sigma(\bfk_1,\bfk_3,\bfk_2) =B_\sigma(-\bfk_1,-\bfk_2,-\bfk_3) $, we can rewrite eq.(\ref{Aterm}) as 
 \begin{equation}
 A(k,\mu)=  -(k \mu) \int d^3 \boldsymbol{k'} \left[  \frac{k_z '}{k'^2} B_\sigma(\boldsymbol{k'},\boldsymbol{k}-\boldsymbol{k'},-\boldsymbol{k}) +  \frac{k\mu-k_z'}{|\bfk-\bfk'^2|} B_\sigma(\boldsymbol{k}-\boldsymbol{k'}, \boldsymbol{k'},-\boldsymbol{k}) \right]
 \label{Aterm2}
 \end{equation}
$F(\boldsymbol{k})$ is already 2nd order in the linear power spectrum. In terms of the perturbation kernels we can write it as 
 \begin{equation}
 F(\boldsymbol{k}) = \frac{k_z}{k^2}G_1(k) \left[ F_1(k) P_0 (k) - \frac{k_z^2}{k^2} G_1(k)P_0 (k) \right]
 \end{equation}
 The main feature of this model is the inclusion of the $A$ and $B$ correction terms which account for higher-order interactions between the density and velocity fields. This gives the model good predictive power at weakly nonlinear scales, as shown by $N$-body comparisons \cite{Taruya:2010mx,Nishimichi:2011jm,Taruya:2013my}. In GR these terms have been shown to enhance the power spectrum amplitude at the BAO scale and have a non-negligible effect on the shape of the power spectrum \cite{Taruya:2010mx}.  In modified gravity theories the $B$ term is generally expected to be enhanced because of its linear growth dependance while the $A$ term involves the 2nd order perturbations so it is not obvious how it will change. Both do require correct modelling within your gravitational model's context as shown in Ref.\cite{Taruya:2013quf}. We also note that so far no dynamics for the perturbations have been taken into account and so the expressions in this section are valid for the considered  class of modified gravity models. 
\newline
\newline
Eq.(\ref{redshiftps}) gives a non-linear prediction for the redshift space power spectrum. In GR this expression has been computed up to 2-loop order in the resummed PT scheme \cite{Taruya:2013my,Taruya:2009ir,Okamura:2011nu,Crocce:2007dt,Crocce:2012fa,Taruya:2012ut} which has a larger range of validity over the SPT treatment. Despite this, as well as the convergence problems of the SPT treatment, the scheme still gives us a good working range of scales in the quasi non-linear regime making it well suited for the goal of probing gravity.   
\newline
\newline
In the next section we detail a tool for calculating the observables discussed in this section and present some results produced by the code. These include comparisons of the numerical results with the analytic expressions for 3 different cosmologies as well as a reproduction of some of the results of Ref.\cite{Taruya:2013quf}. 


\section{Results}
\subsection{Code and Procedure} 
For our purposes we seek to calculate the density and velocity perturbations up to 3rd order which will be used in the computation of the 1-loop power spectrum as well as the $A$ correction term up to consistent order. To this end, we require the, generally scale and time dependent, perturbative kernels $F_1(\bfk ; a)$, $G_1(\bfk; a)$, $F_2(\bfk_1,\bfk_2; a)$, $G_2(\bfk_1,\bfk_2; a)$, $F_3(\bfk_1,\bfk_2,\bfk_3; a)$ and $G_3(\bfk_1,\bfk_2,\bfk_3; a)$. We use the scale factor $a$ as our time variable.   Also recall that under the integration constraints (eqs.(\ref{eq:psconstraint1}), (\ref{eq:psconstraint0})), the specific kernels we are after are those in eq.(\ref{eq:loopkernels}), $F_1(k ;a)$, $G_1(k;a), F_2(k,r,x;a)$, $G_2(k,r,x;a)$, $F_3(k,r,x;a)$ and $G_3(k,r,x;a)$ which denote the 1st, 2nd and 3rd order kernels respectively. 
\newline
\newline
It is well known that for the LCDM model the perturbations evolve independently of scale at linear order and so separability of the kernels into  time and  scale dependent parts becomes a good assumption allowing for an analytic solution. This is true for those models in which  $\mu(k;a)$, $\gamma_2(\bfk, \bfk_1, \bfk_2; a)$  and $\gamma_3(\bfk, \bfk_1, \bfk_2, \bfk_3;a)$ can be written as separable functions of scale and time. Analytic forms of the perturbations  for massless Horndeski's theory under the quasi-static approximation have been derived in Refs.\cite{Takushima:2015iha,Takushima:2013foa,Kimura:2011dc,DeFelice:2011hq}, going up to 3rd order as well as forms for the 1-loop power spectra in Ref.\cite{Takushima:2015iha}, and in Ref.\cite{Koyama:2009me} the authors derive the analytical expressions for the perturbations up to 3rd order in 5-dimensional Dvali-Gabadadze-Porratti (DGP) gravity \cite{Dvali:2000hr} as well as the 1-loop power spectra. 
\newline
\newline
In general separability cannot be assumed, for example in the well known $f(R)$ class of models (see Ref.\cite{Capozziello:2007ec} and Ref.\cite{Sotiriou:2008it} for reviews) . In this case the Euler and continuity equations become analytically intractable and one must calculate the kernels numerically to proceed. Since our goal is to be as general as possible,  we solve eq.\ref{eq:dynamicsystem} numerically. We refer the readers to Ref.\cite{Taruya:paper} and Ref.\cite{Taruya:talk} for details on the numerical algorithm. The algorithm involves finding kernel solutions for various values of integrated Fourier vector magnitude $x$ and angular parameter $\mu$. We denote the number of $\mu$ values  we sample by $n_1$ and the number of $x$ values by $n_2$ giving a total of $n_1\times n_2$ solutions. Increasing these numbers gives a finer solution space to sample and consequently a truer result for our observables.  In general we also need to initialise the $n_1 \times n_2$ solutions for each $k$  when the source terms of the Poisson equation are $k$ dependent. If this is not the case we can initialise the kernels once.  
\newline
\newline
Our code is written in {\sc c++} and is based off the cosmological perturbation theory code, Copter \cite{copter}. The Euler and continuity system is solved using the $gsl$ package {\bf odeiv2} . {\bf odeiv2} is able to solve the equations using a host of methods. The default for our code is a Runge-Kutta Prince-Dormand (8, 9) method which works very well and quickly. The number of time steps used by the solver is adaptive and depends on the desired accuracy. Our default accuracy is based on matching the numerically calculated $P_{22}$ and $P_{13}$ with the Einstein-de Sitter (FRW with $\Omega_m=1$) analytical versions to below 1 percent within the desired range of scales. For the numerically calculated kernels we use Einstein-de Sitter (EDS) initial conditions in our system of equations, as in most dark energy and gravity models the EDS approximations hold at early times 
\begin{align}
F_1(\bfk;a_0) &= a_0  \\ 
G_1(\bfk; a_0) &= -a_0 \\ 
F_2(\bfk_1,\bfk_2; a_0) &= a_0^2 F_2^{sym}(\bfk_1,\bfk_2) \\ 
G_2(\bfk_1,\bfk_2;a_0) &= a_0^2 G_2^{sym}(\bfk_1,\bfk_2) \\ 
F_3(\bfk_1,\bfk_2,\bfk_3; a_0)& = a_0^3 F_3^{sym}(\bfk_1,\bfk_2,\bfk_3) \\ 
G_3(\bfk_1,\bfk_2,\bfk_3; a_0)& = -a_0^3 G_3^{sym}(\bfk_1,\bfk_2,\bfk_3) \\ 
\end{align}
where the symmetric analytic kernels on the right hand side are given in  Appendix A. The complete system of Euler and continuity equations to be solved consists of 9 pairs.  
\begin{itemize}
\item
3 first order set which is solved for $F_1(k_i; a)$ and $G_1(k_i;a)$, where $k_i \in \{k,k',|k-p|\}$. 
\item 
5 second order sets which are solved for  $F_2(\bfk_1, \bfk_2; a)$ and $G_2(\bfk_1, \bfk_2;a)$. The 5 sets are for the following pairs of $k$-mode inputs : $(\bfk', \bfk-\bfk')$, $(\bfk-\bfk', -\bfk)$, $(-\bfk', \bfk)$, $(\bfk', \bfk)$ and $(\bfk', -\bfk')$.  The first of these is for $P_{22}$ while the last 4 are used to construct the 3rd order kernel and the RSD A-term. 
\item 
1 third order set which is solved for  $F_3(\bfk, \bfk', -\bfk' ; a)$ and $G_3(\bfk, \bfk', -\bfk' _1;a)$
\end{itemize}
To determine the realm of validity for the SPT calculations and hence the sampling size, we solve the following equation 
\begin{equation}
\frac{k^2}{6\pi^2} \int^{k_{max}}_0 dq P_L(q;a) = 0.18
\label{validityrange}
\end{equation}
This range is the 1 \% accuracy regime, found empirically by comparing perturbation theory with N-body simulations for GR \cite{Nishimichi:2008ry}. It is useful in providing a rough realm of validity in the general model case. For our convergence tests we plot up to a $k_{max} = 0.2$ which was calculated using the above relation in a LCDM cosmology for $z = 0.4$  which is the upper redshift on DESI's bright galaxy survey \cite{DESI,Levi:2013gra}, the lowest redshift survey it will undertake. It will also look at 18 million emission line galaxies in the redshift range $0.6 \leq z \leq 1.6$. Perturbation theory is known to do better at higher redshift (for example Ref.\cite{Carlson:2009it}), giving a larger realm of validity for this and other upcoming surveys. We also note here than the validity range ($k_{max}$) is expected to be smaller for the nDGP case because of the fifth force's enhancement of velocities and clustering at larger scales. This holds true for most modified models of gravity.  
\newline
\newline
Next we present 3 checks of the code against analytic results and a comparison of its predictions against N-body results in $f(R)$ gravity. 

\subsection{Performance of Numerical Algorithm: Comparing to Analytic Forms}
\begin{figure}[H]
  \captionsetup[subfigure]{labelformat=empty}
  \centering
  \subfloat[]{\includegraphics[width=8.3cm, height=8.3cm]{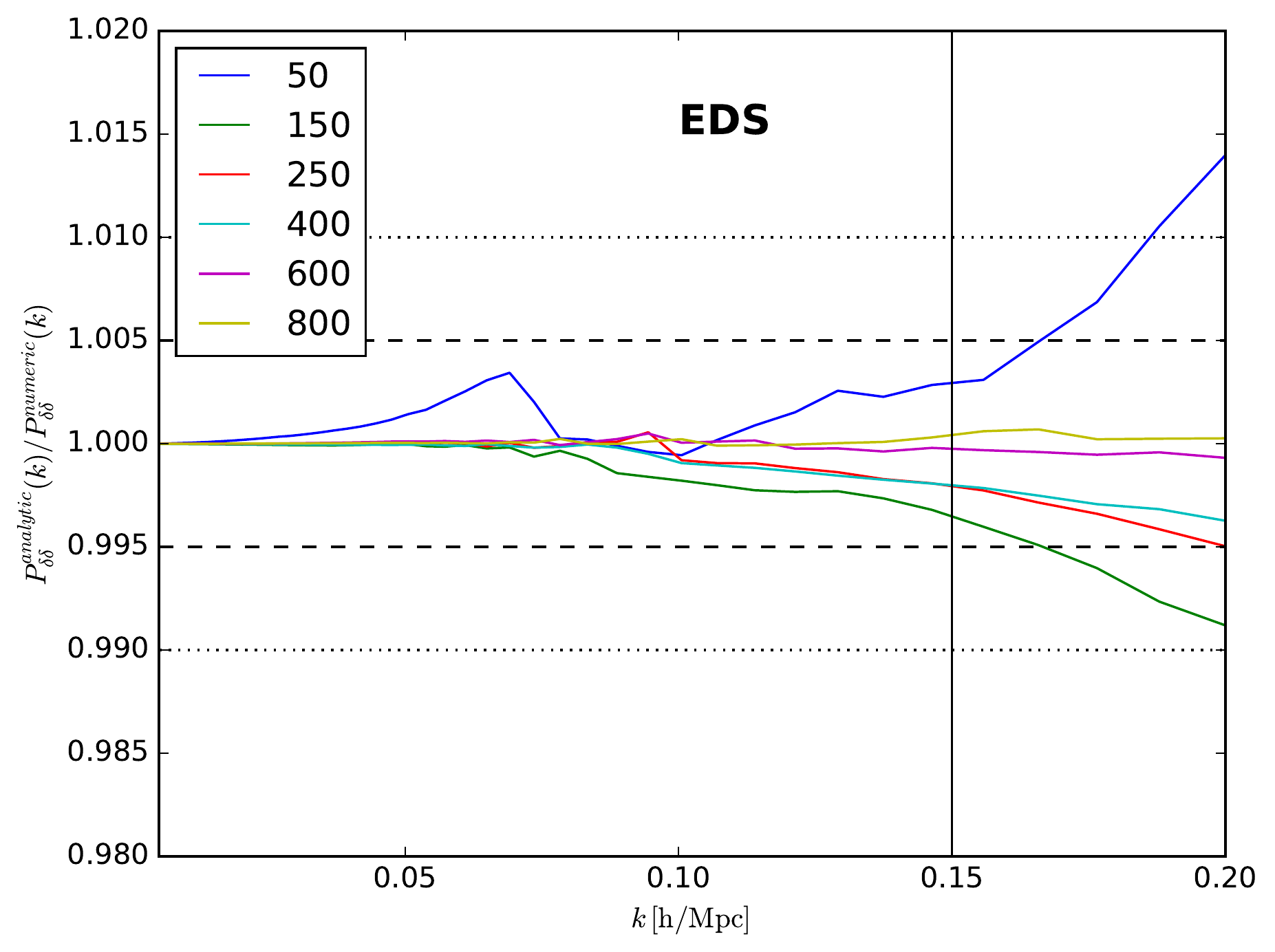}} \quad
  \subfloat[]{\includegraphics[width=8.3cm, height=8.3cm]{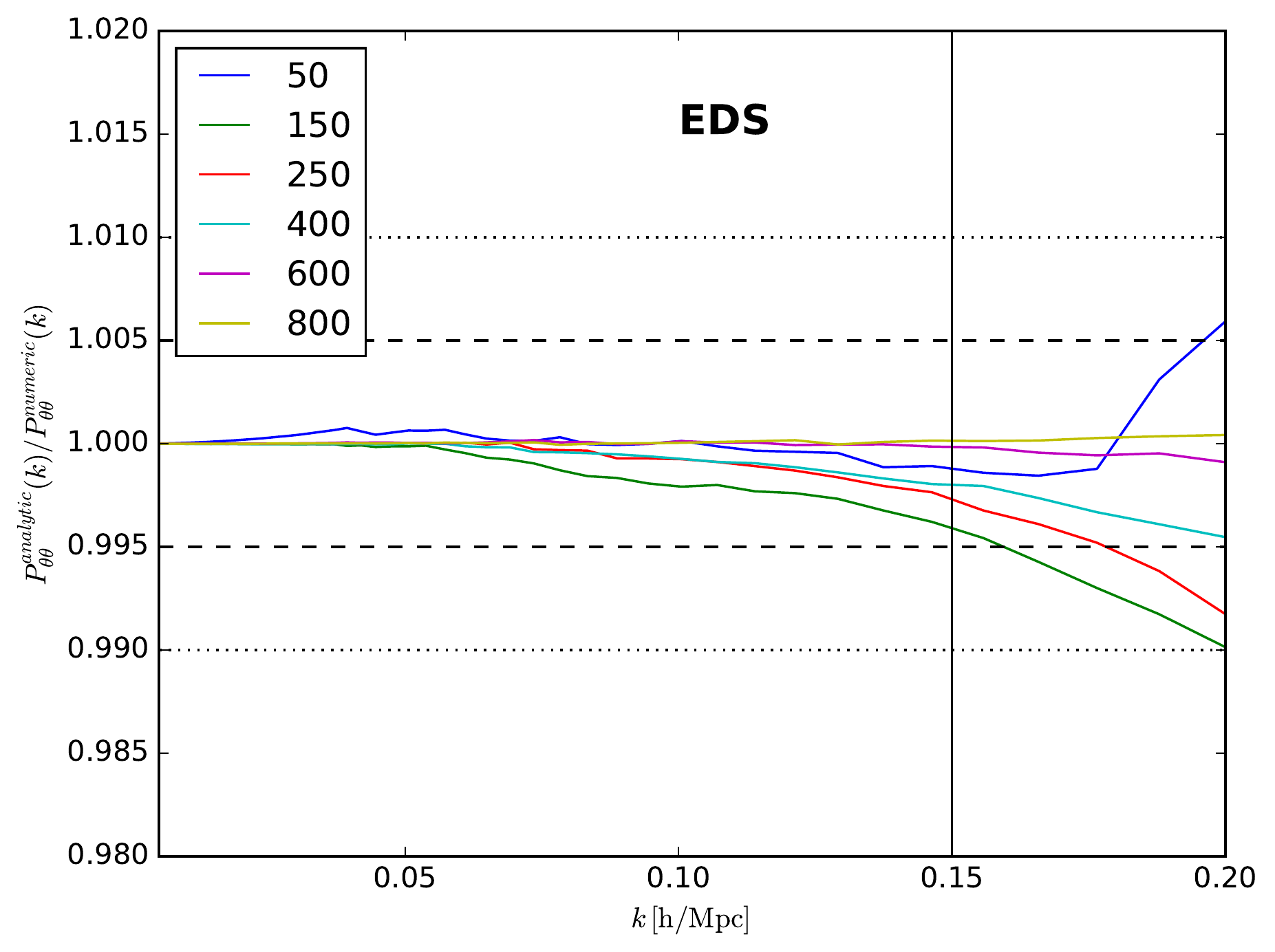}} 
  \caption[CONVERGENCE ]{Test for convergence of the numerical to analytical 1-loop matter  (left) and velocity (right) power spectrum in the Einstein-de Sitter cosmology for $n1=n2=50,150,250,400$ and $800$.}
\label{convergence1}
\end{figure}

\begin{figure}[H]
  \captionsetup[subfigure]{labelformat=empty}
  \centering
  \subfloat[]{\includegraphics[width=8.3cm, height=8.3cm]{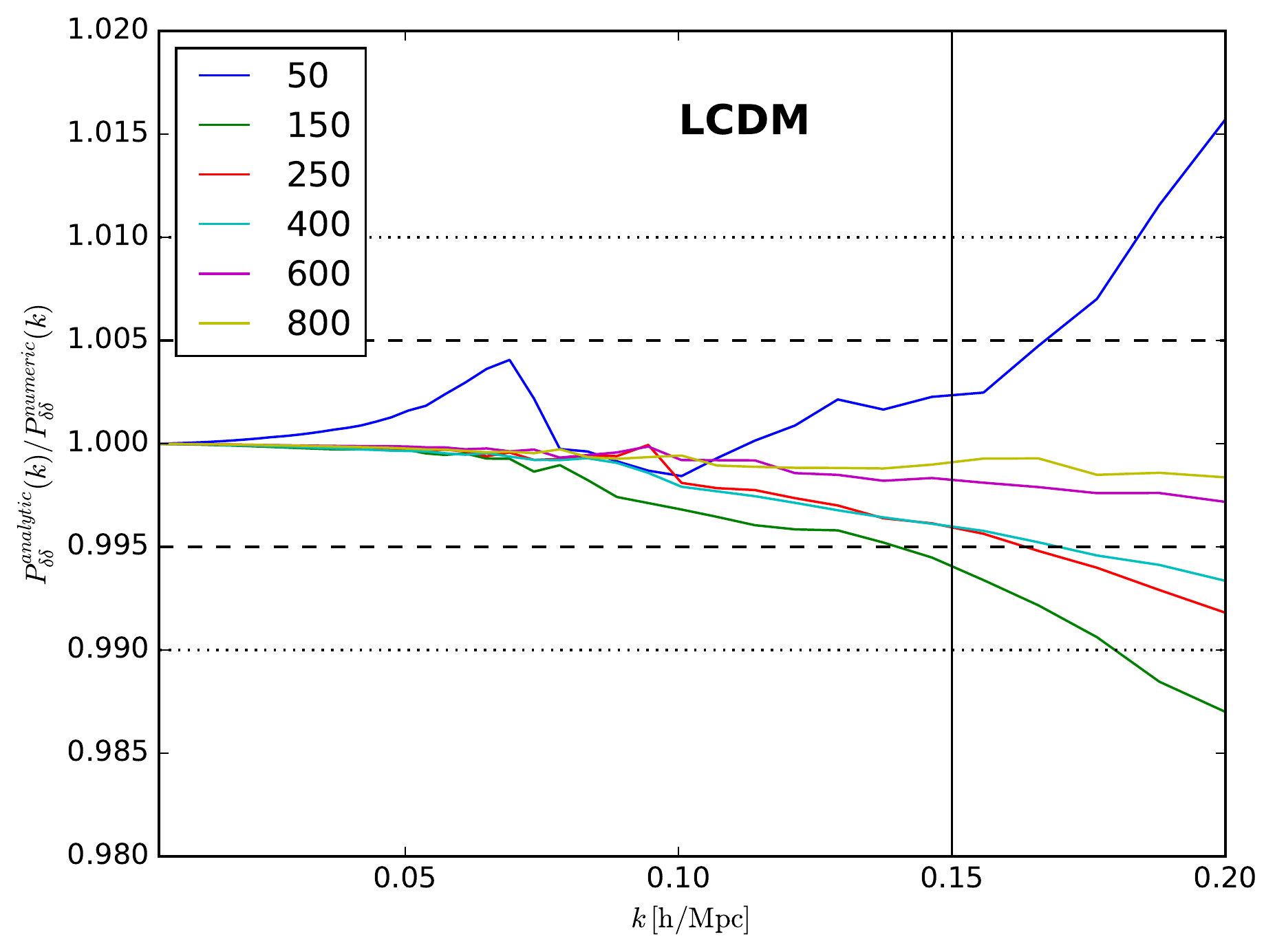}} \quad
  \subfloat[]{\includegraphics[width=8.3cm, height=8.3cm]{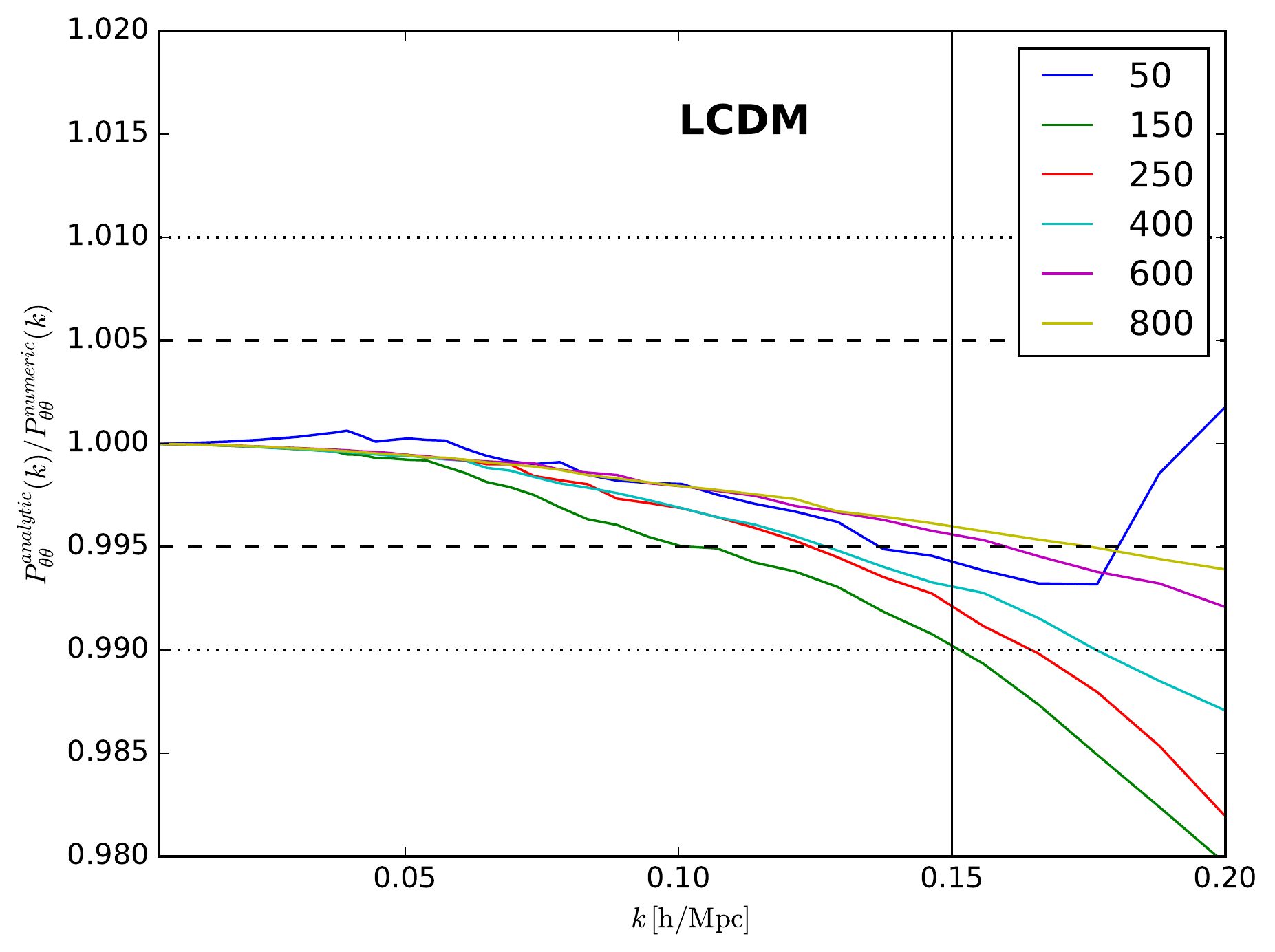}} 
  \caption[CONVERGENCE ]{Test for convergence of the numerical to analytical 1-loop matter  (left) and velocity (right) power spectrum   in the LCDM cosmology for $n1=n2=50,150,250,400$ and $800$.}
\label{convergence2}
\end{figure}

\begin{figure}[H]
  \captionsetup[subfigure]{labelformat=empty}
  \centering
  \subfloat[]{\includegraphics[width=8.3cm, height=8.3cm]{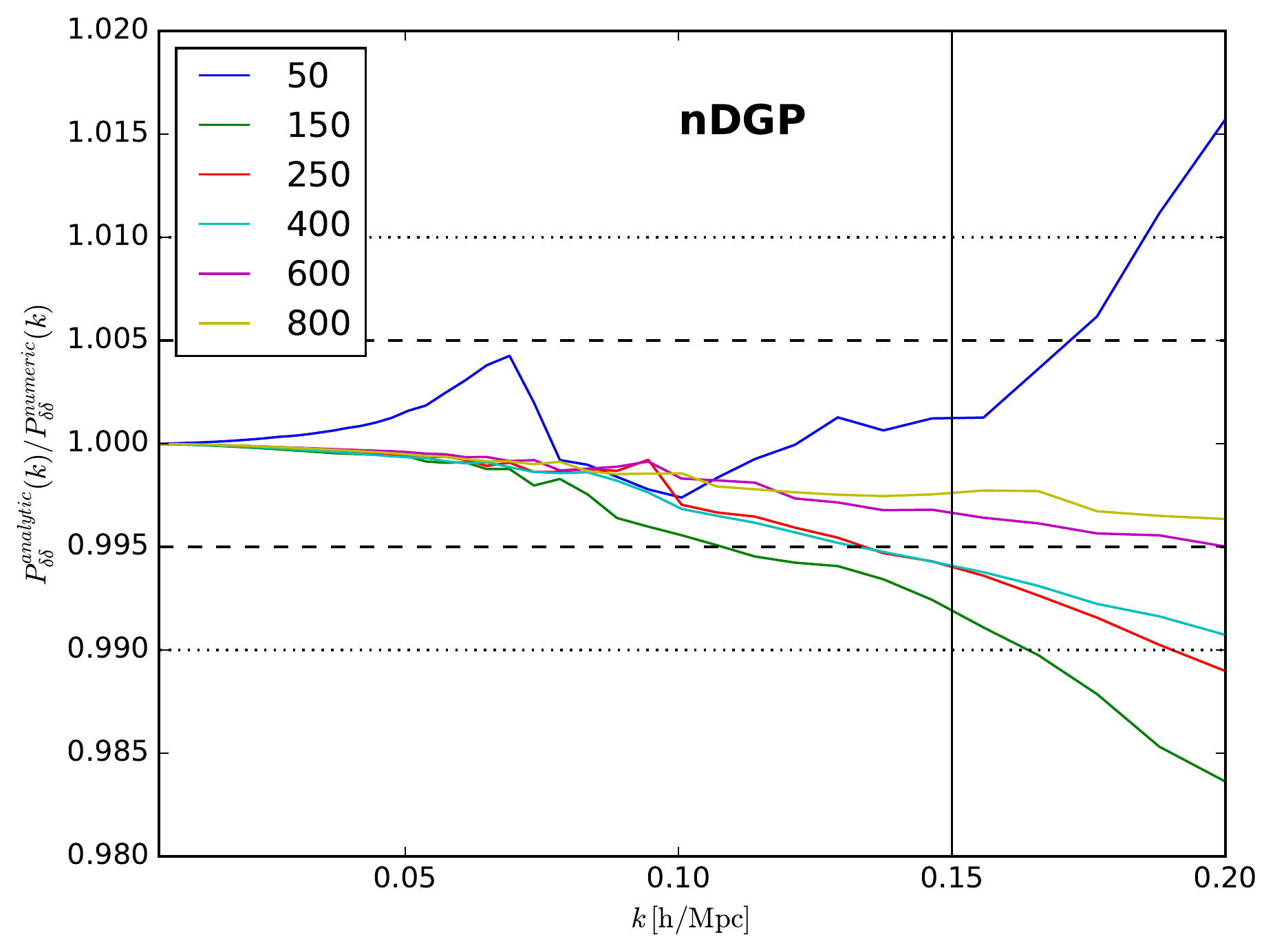}} \quad
  \subfloat[]{\includegraphics[width=8.3cm, height=8.3cm]{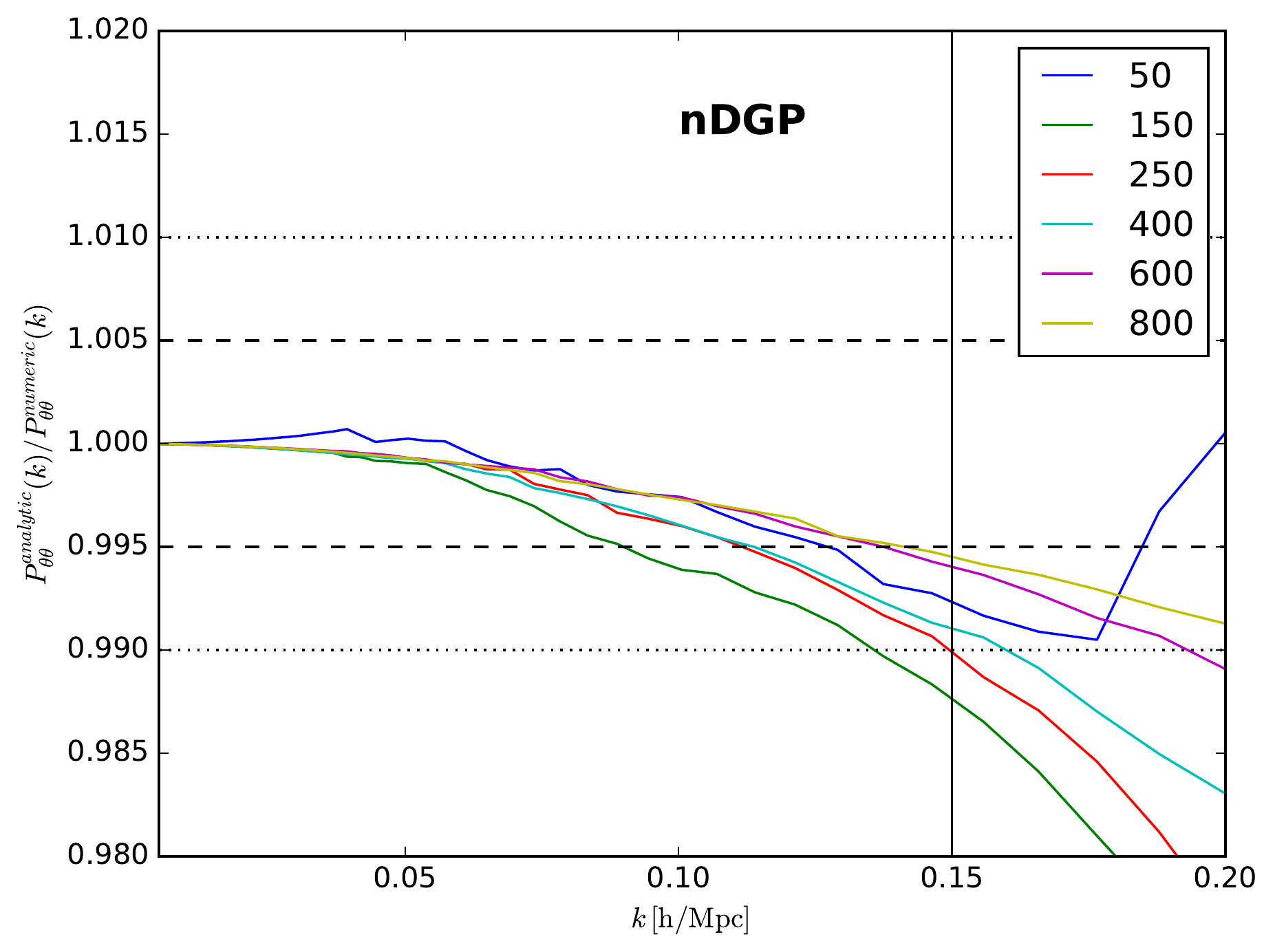}} 
  \caption[CONVERGENCE ]{Test for convergence of the numerical to analytical 1-loop matter  (left) and velocity (right) power spectrum   in nDGP gravity for $n1=n2=50,150,250,400$ and $800$.}
\label{convergence3}
\end{figure}

\begin{figure}[H]
  \captionsetup[subfigure]{labelformat=empty}
  \centering
  \subfloat[]{\includegraphics[width=8.3cm, height=8.3cm]{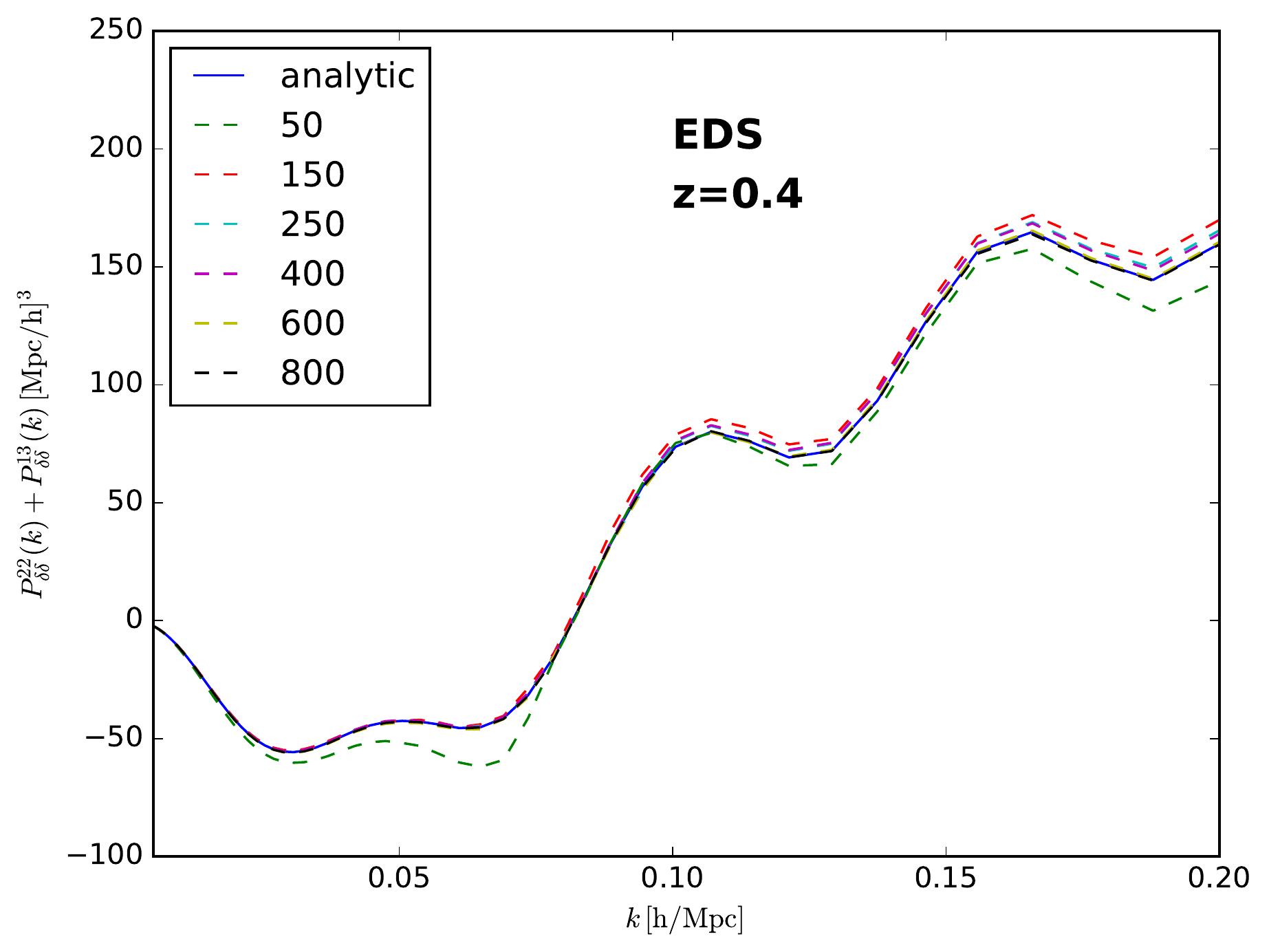}} \quad
  \subfloat[]{\includegraphics[width=8.3cm, height=8.3cm]{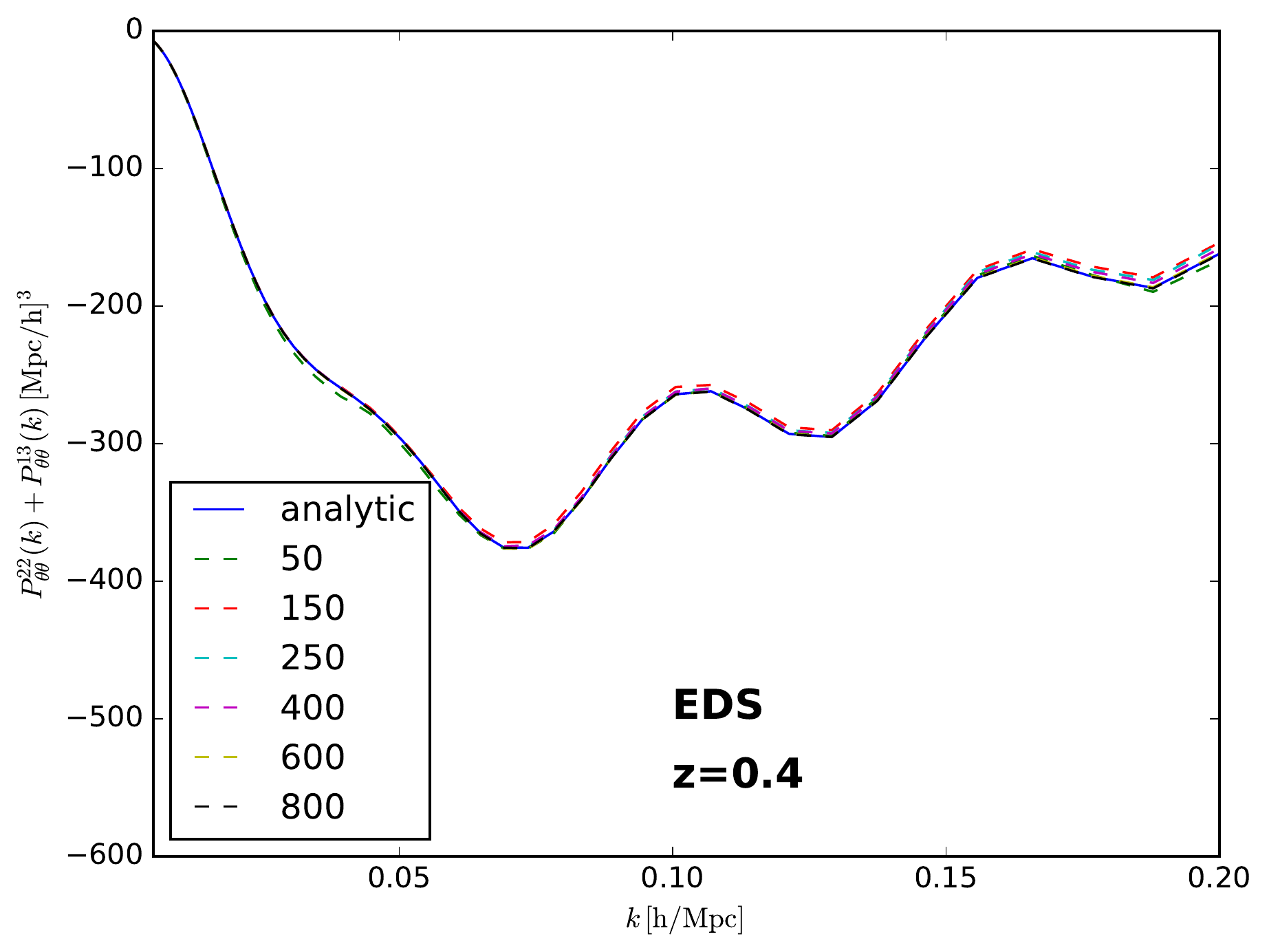}} 
  \caption[CONVERGENCE ]{Test for convergence of the numerical to analytical matter  (left) and velocity (right) 1-loop contributions in the Einstein-de Sitter cosmology for $n1=n2=50,150,250,400,600$ and $800$.}
\label{convergence4}
\end{figure}

\begin{figure}[H]
  \captionsetup[subfigure]{labelformat=empty}
  \centering
  \subfloat[]{\includegraphics[width=8.3cm, height=8.3cm]{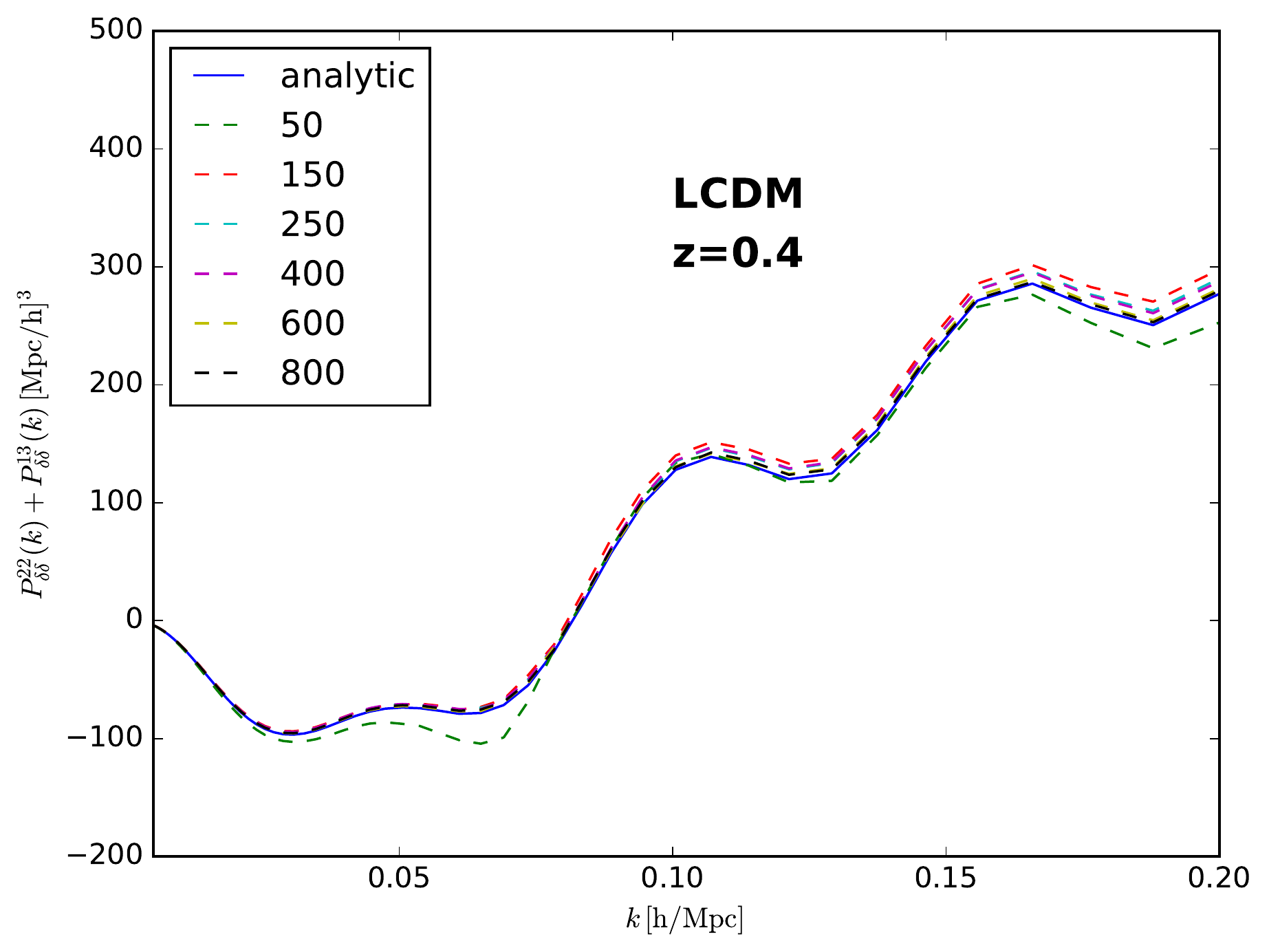}} \quad
  \subfloat[]{\includegraphics[width=8.3cm, height=8.3cm]{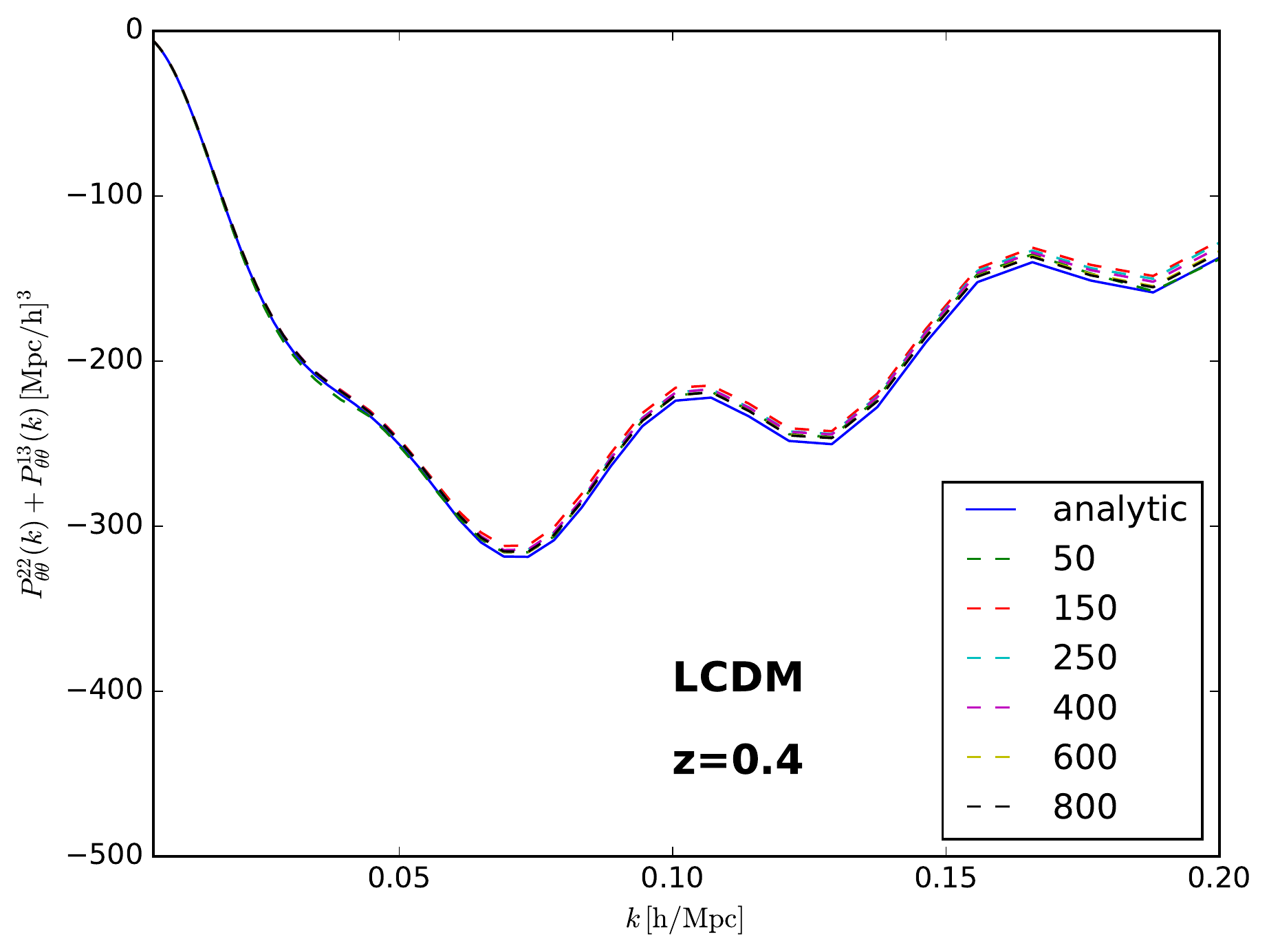}} 
  \caption[CONVERGENCE ]{Test for convergence of the numerical to analytical matter (left) and velocity  (right) 1-loop contributions the LCDM cosmology for $n1=n2=50,150,250,400,600$ and $800$.}
\label{convergence5}
\end{figure}

\begin{figure}[H]
  \captionsetup[subfigure]{labelformat=empty}
  \centering
  \subfloat[]{\includegraphics[width=8.3cm, height=8.3cm]{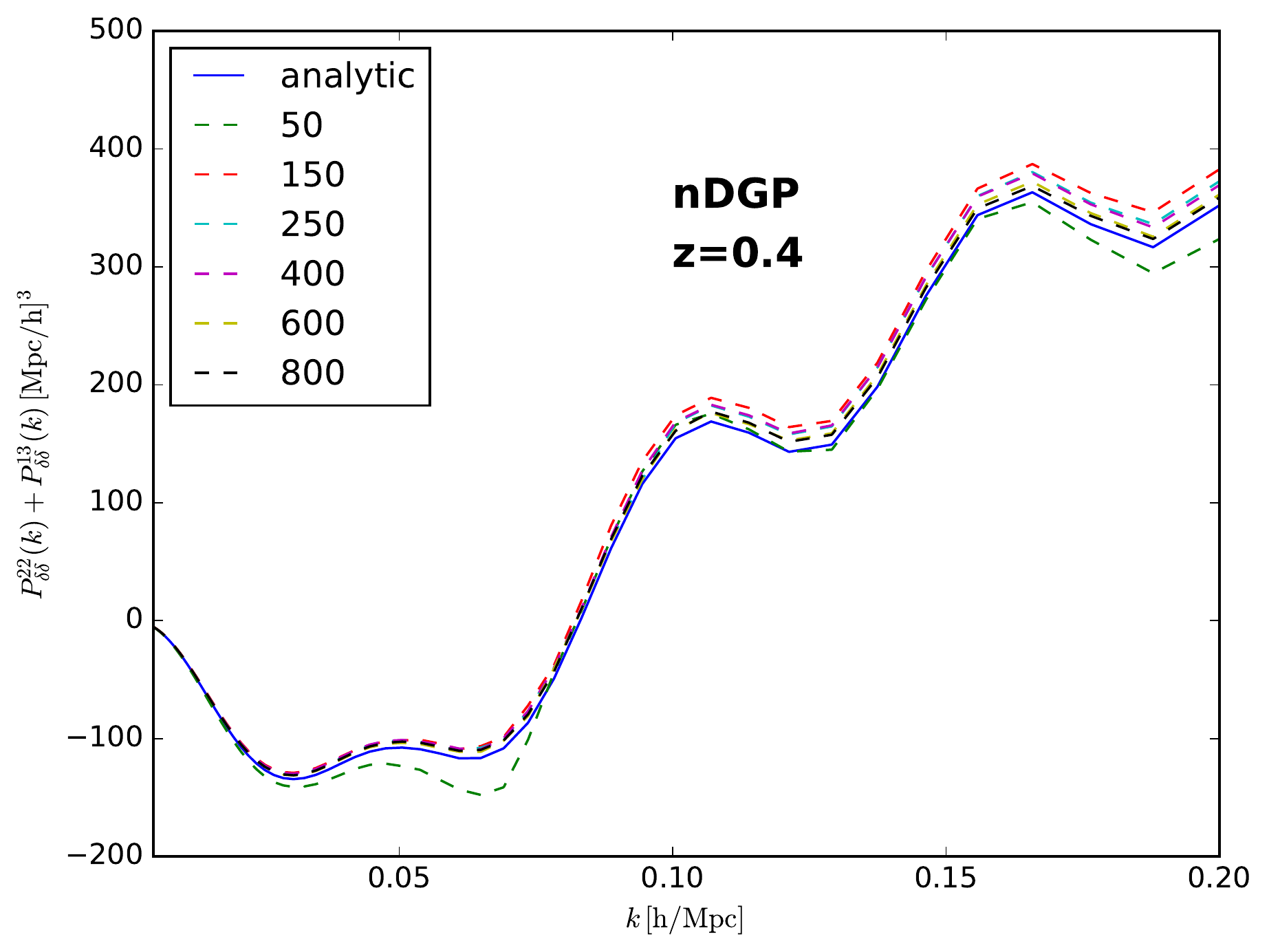}} \quad
  \subfloat[]{\includegraphics[width=8.3cm, height=8.3cm]{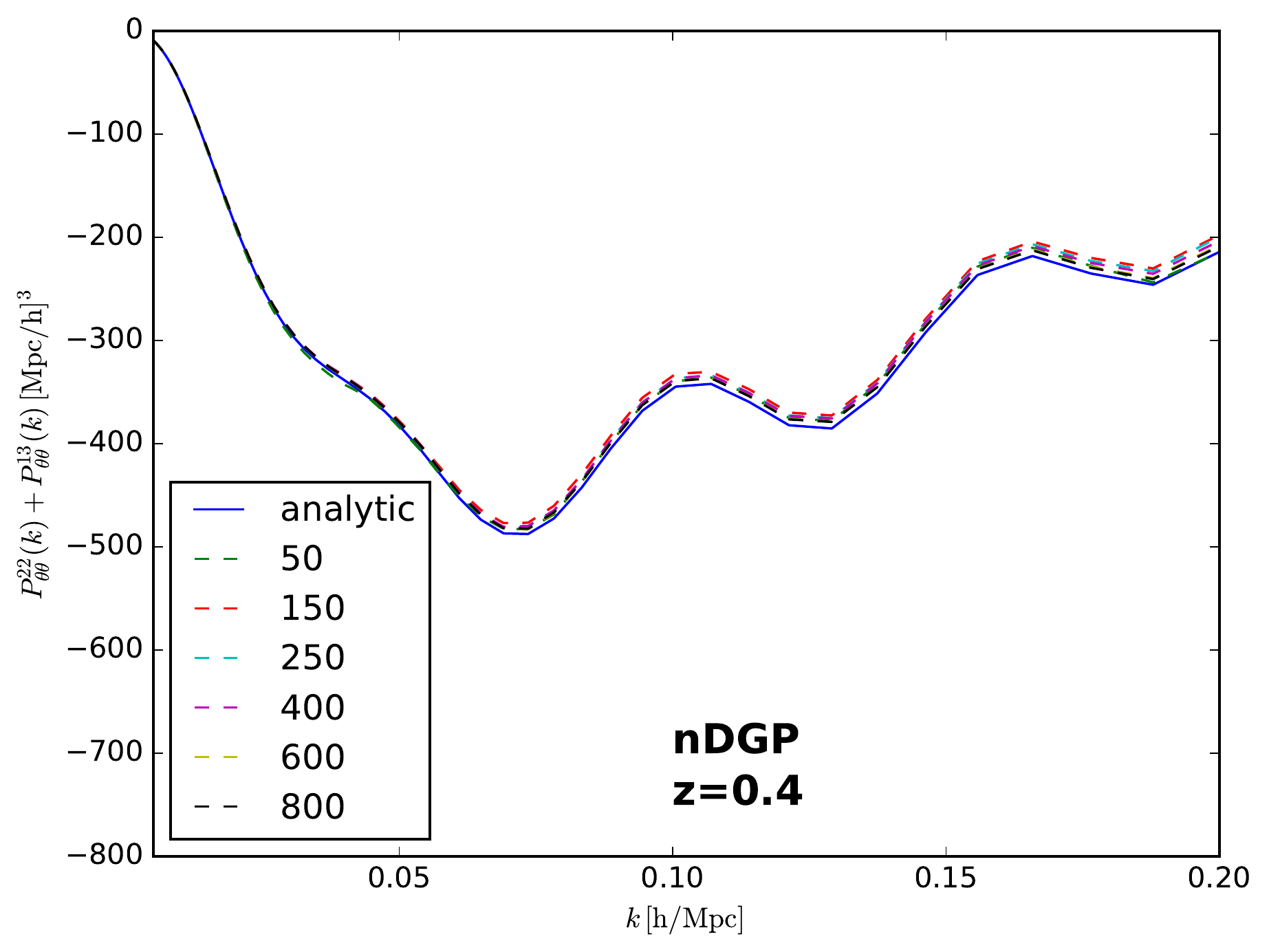}} 
  \caption[CONVERGENCE ]{Test for convergence of the numerical to analytical matter (left) and velocity (right)1-loop contributions in nDGP gravity for $n1=n2=50,150,250,400,600$ and $800$.}
\label{convergence6}
\end{figure}

\begin{figure}[H]
  \captionsetup[subfigure]{labelformat=empty}
  \centering
  \subfloat[]{\includegraphics[width=10.5cm, height=8.3cm]{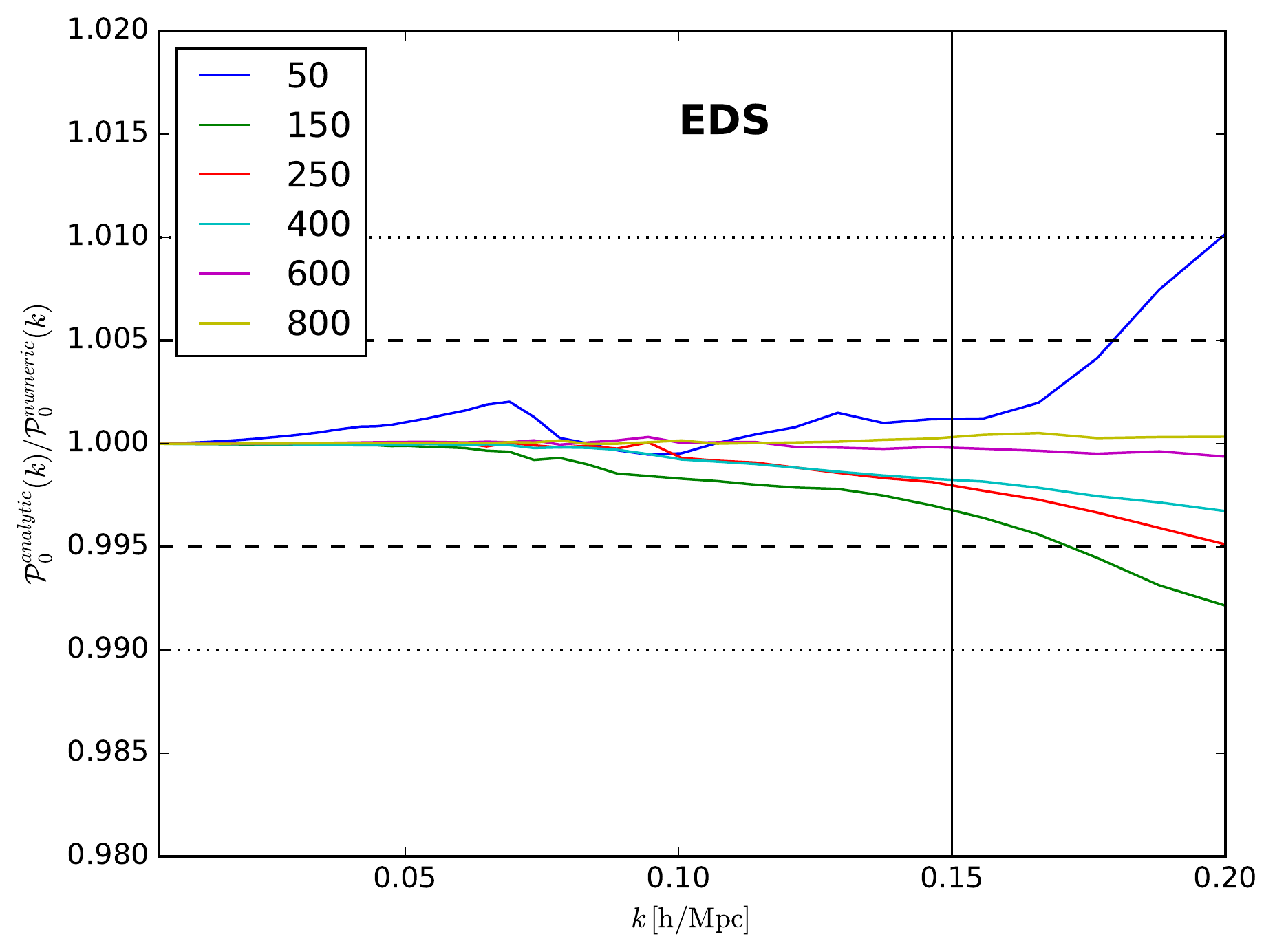}} 
  \caption[CONVERGENCE ]{Test for convergence of the numerical to analytical TNS redshift space monopole of the power spectrum in the Einstein-de Sitter cosmology for $n1=n2=50,150,250,400$ and $800$}
\label{convergence7}
\end{figure}

\begin{figure}[H]
  \captionsetup[subfigure]{labelformat=empty}
  \centering
  \subfloat[]{\includegraphics[width=8.3cm, height=8.3cm]{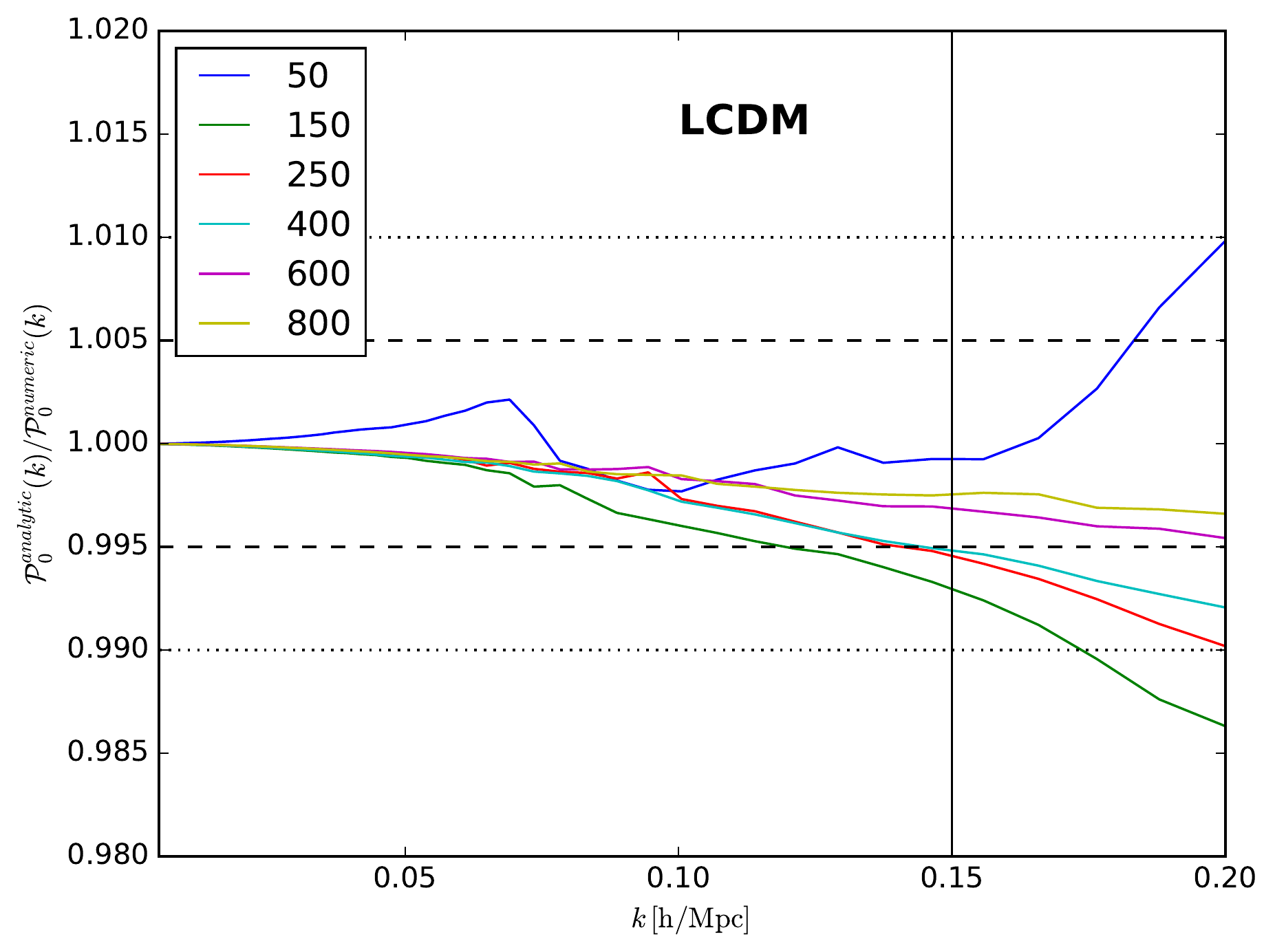}} \quad
  \subfloat[]{\includegraphics[width=8.3cm, height=8.3cm]{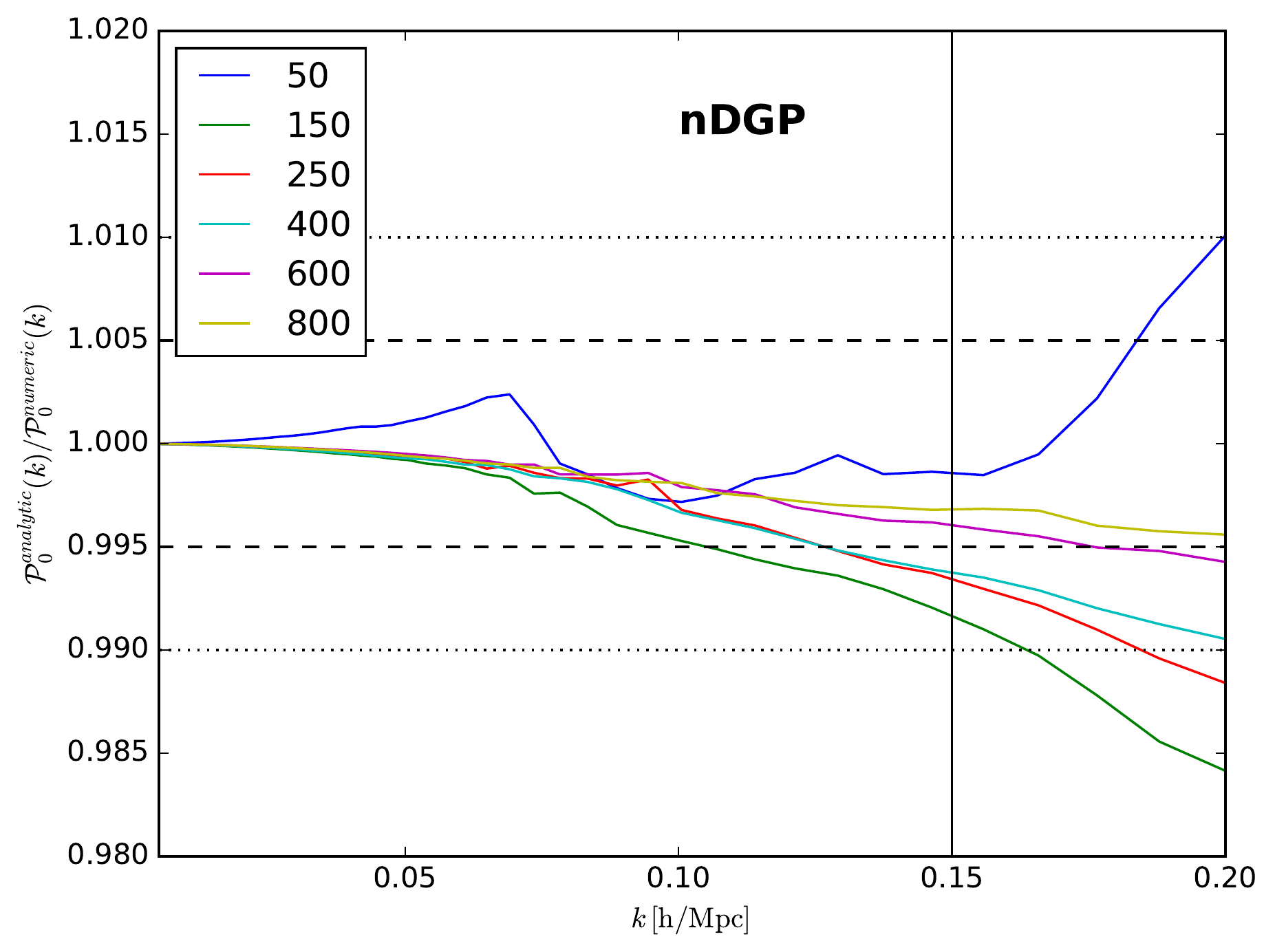}} 
  \caption[CONVERGENCE ]{Test for convergence of the numerical to analytical TNS redshift space monopole of the power spectrum in the LCDM cosmology(left) and in nDGP gravity(right) for $n1=n2=50,150,250,400$ and $800$}
\label{convergence8}
\end{figure}

\begin{figure}[H]
  \captionsetup[subfigure]{labelformat=empty}
  \centering
  \subfloat[]{\includegraphics[width=8.3cm, height=8.3cm]{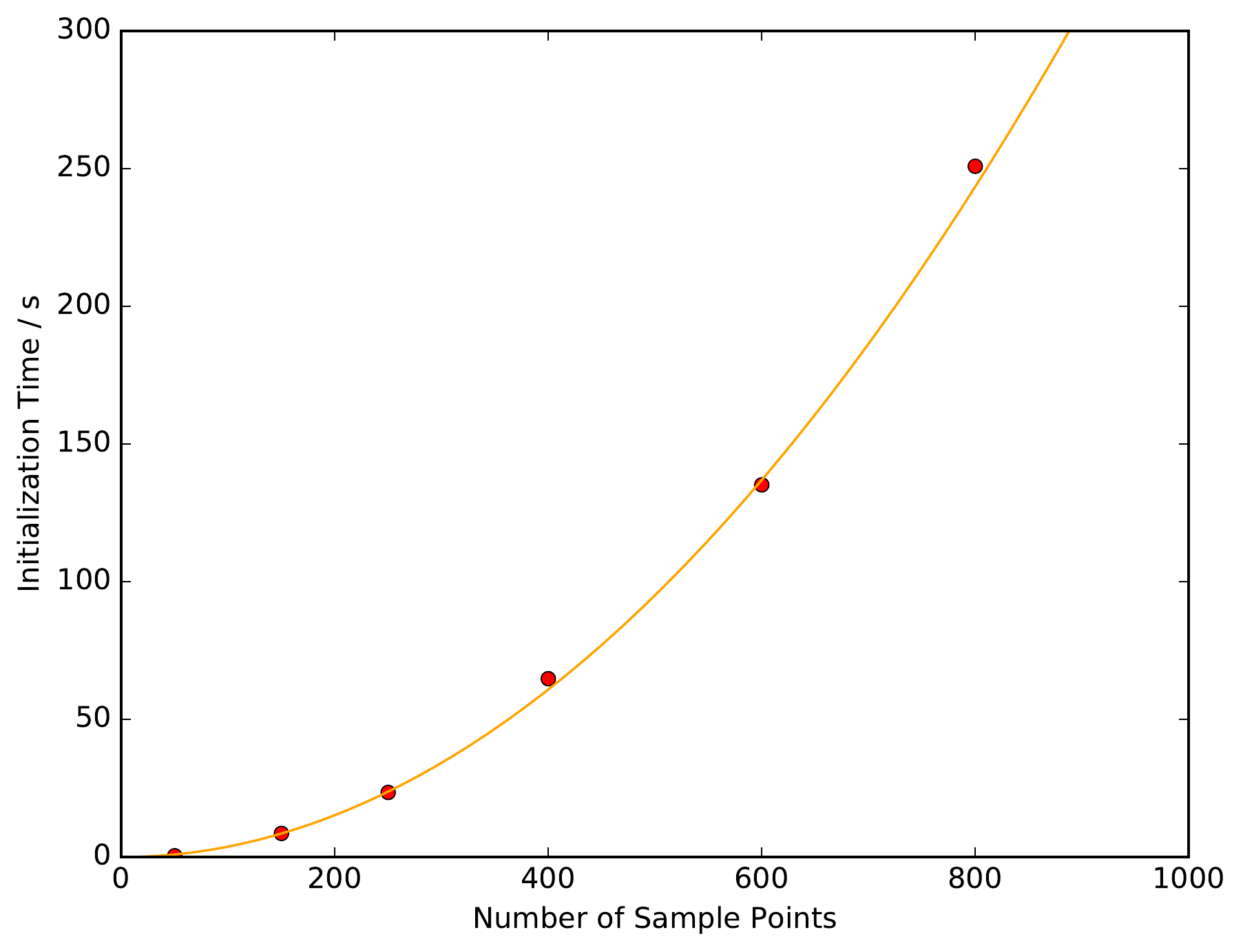}} \quad
  \subfloat[]{\includegraphics[width=8.3cm, height=8.3cm]{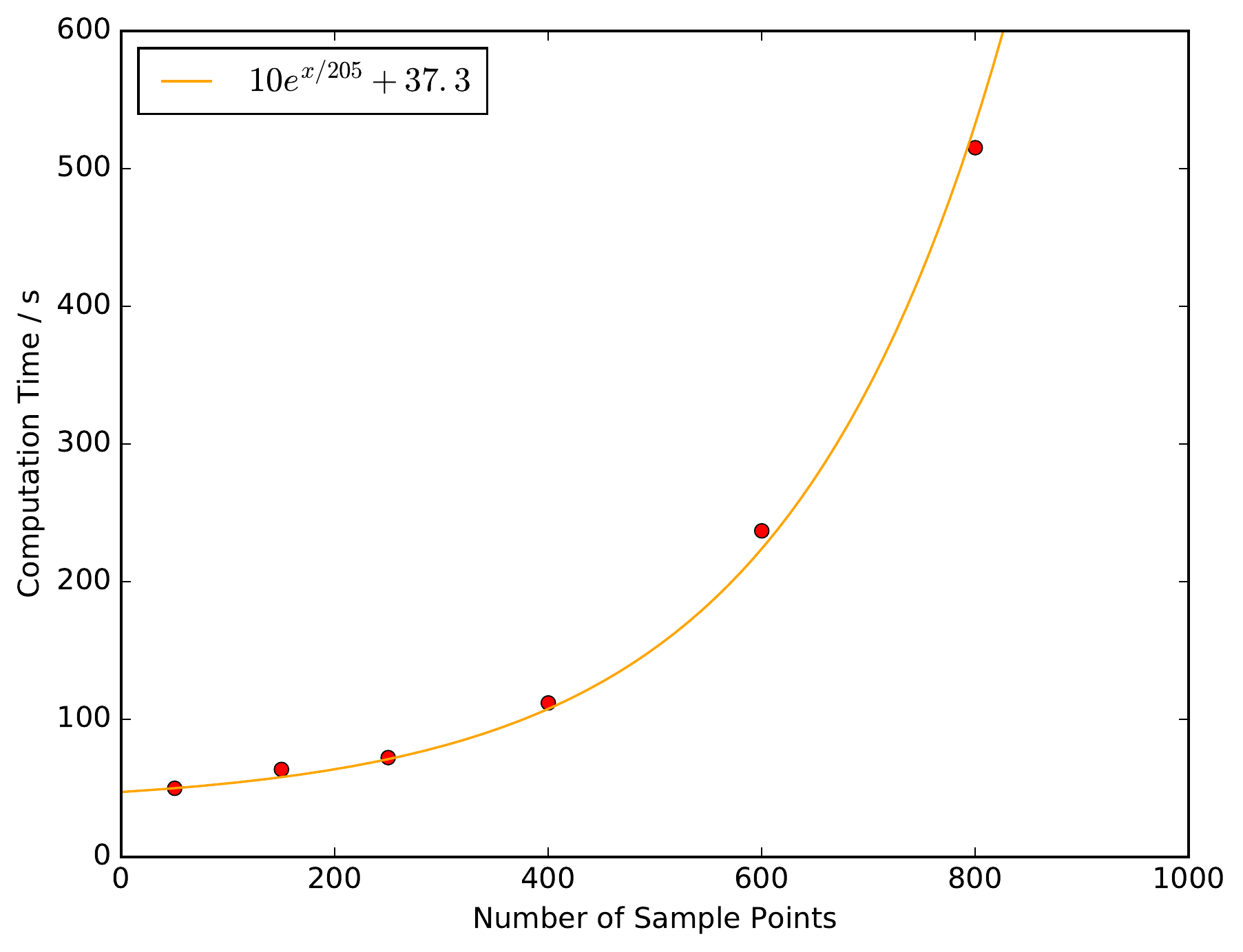}} 
  \caption[CONVERGENCE ]{Time taken to initialise the kernels for different values of $n1=n2$ (left) and the time taken to compute 30 $k$ modes of the 1-loop power spectrum for different values of $n1=n2$ for a single initialisation (right). Results are for the nDGP case. Quadratic (left) and exponential (right) curves have been fit to the points.}
\label{time1}
\end{figure}

Here we discuss the performance of the code  in reproducing analytic results. In this section all results are based on a primordial linear power spectrum generated by the Boltzman solver CLASS \cite{Blas:2011rf} with a flat LCDM cosmology with the following parameters $h=0.697$, $n=0.971$, $\Omega_b= 0.046$, $\Omega_m = 0.281$ and  $\sigma_8=0.82$. All timing results were obtained on a MacBook Pro laptop computer, with a 2.52 GHz Intel Core 2 Duo processor and running on Mac OS X version 10.6.8.  The code is both OpenMP and MPI enabled but  parallelisation was not used for the timing results. 
\newline
\newline
We take $z=0.4$ corresponding to $a=0.71$, with normalisation $a=1$ at present. We compute the 1-loop correction terms $P^{22}(k)$ and $P^{13}(k)$ as well as the monopole $P_0(k)$ for 30 $k$ modes  taken logarithmically from $k=0.005$h/Mpc to $k=0.2$ h/Mpc, with the power spectrum multipoles being defined by 
\begin{equation}
P_l^{(S)}(k)=\frac{2l+1}{2}\int^1_{-1}d\mu P_{TNS}^{(S)}(k,\mu)\mathcal{P}_l(\mu)
\end{equation}
where $\mathcal{P}_l(\mu)$ denote the Legendre polynomials and $P_{TNS}^{(S)}(k)$ is given by eq.(\ref{redshiftps}). This is done for 3 different cosmologies; the LCDM cosmology describing our linear power spectrum, an Einstein-de Sitter cosmology and for the normal branch of the DGP model \cite{Dvali:2000hr} (nDGP). 
\newline
\newline
The EDS cosmology has a well known separable solution which is also known to be a very good approximation for the LCDM case. The nDGP model is a Vainshtein screened model whose perturbations up to 3rd order have been derived in Ref.\cite{Koyama:2009me} in analytic form. For the nDGP case we take our LCDM cosmology as the background and take $\Omega_{rc} = 1/(4 H_0^2 r_c^2)=0.438$. The presence of analytic expressions for these cases make it possible to make a consistency test  of the numerical algorithm. 
\newline
\newline
We expect the numerical algorithm to exactly reproduce the analytic expressions for the EDS cosmology where no scale dependence is involved in the field's evolution. We also expect very close match for both the LCDM and nDGP cosmologies where scale and time separation of the perturbations is known to be a very good approximation.  
\newline
\newline
Fig.\ref{convergence1},Fig.\ref{convergence2} and Fig.\ref{convergence3} illustrate the convergence of the numerical auto 1-loop  power spectra to the analytic results for the EDS, LCDM and nDGP cases respectively as we increase the amount of sampling. The horizontal lines show the $0.5\%$ and $1\%$ deviations while the vertical line at $k=0.15$ h/Mpc denotes a rough realm of validity for the DGP case. These were computed using a single initialisation of the kernel solution space since  the Euler and Continuity equations in these models are $k$ independent.  We also plot the convergence for the 1-loop part of the power spectra ($P_{22}(k)+P_{13}(k)$)in Fig.\ref{convergence4},Fig.\ref{convergence5} and Fig.\ref{convergence6}. Fig.\ref{convergence7} and Fig.\ref{convergence8} show the same convergence for the TNS monopole.
\newline
\newline
Fig.\ref{time1} shows the time cost of finer sampling in the independent case, where the solution space is initialised once. A quadratic curve is fit to the left plot to show the expected trend of cost with sampling since we have $ n1 \times n2 = n1^2$ iterations of the solver. In the right plot we note an exponential increase in the time it takes to compute 30 values of the 1-loop power spectrum as we increase the sampling. This increase probably comes from the added fineness of the solution grid which the power spectrum integral takes values from. 
\newline
\newline
Based on the EDS case we find that $n1=n2=150$ is sufficient sampling to achieve below percent accuracy within the validity range and can be safely used as a standard for the general scale independent case. With this sampling size the numerical algorithm computes 30 1-loop power spectra values  in under a minute. If  further accuracy is required the sampling can easily be increased at a time cost. Parallelisation over $k$ using MPI and Openmp makes it possible to get around this time cost very easily. 
\newline
\newline
In the next section we look at $f(R)$ gravity as a scale dependent test of the code. 
\subsection{Performance of Numerical Algorithm: Scale Dependent Perturbations in $f(R)$}
\begin{figure}[H]
  \captionsetup[subfigure]{labelformat=empty}
  \centering
  \subfloat[]{\includegraphics[width=8.3cm, height=8.3cm]{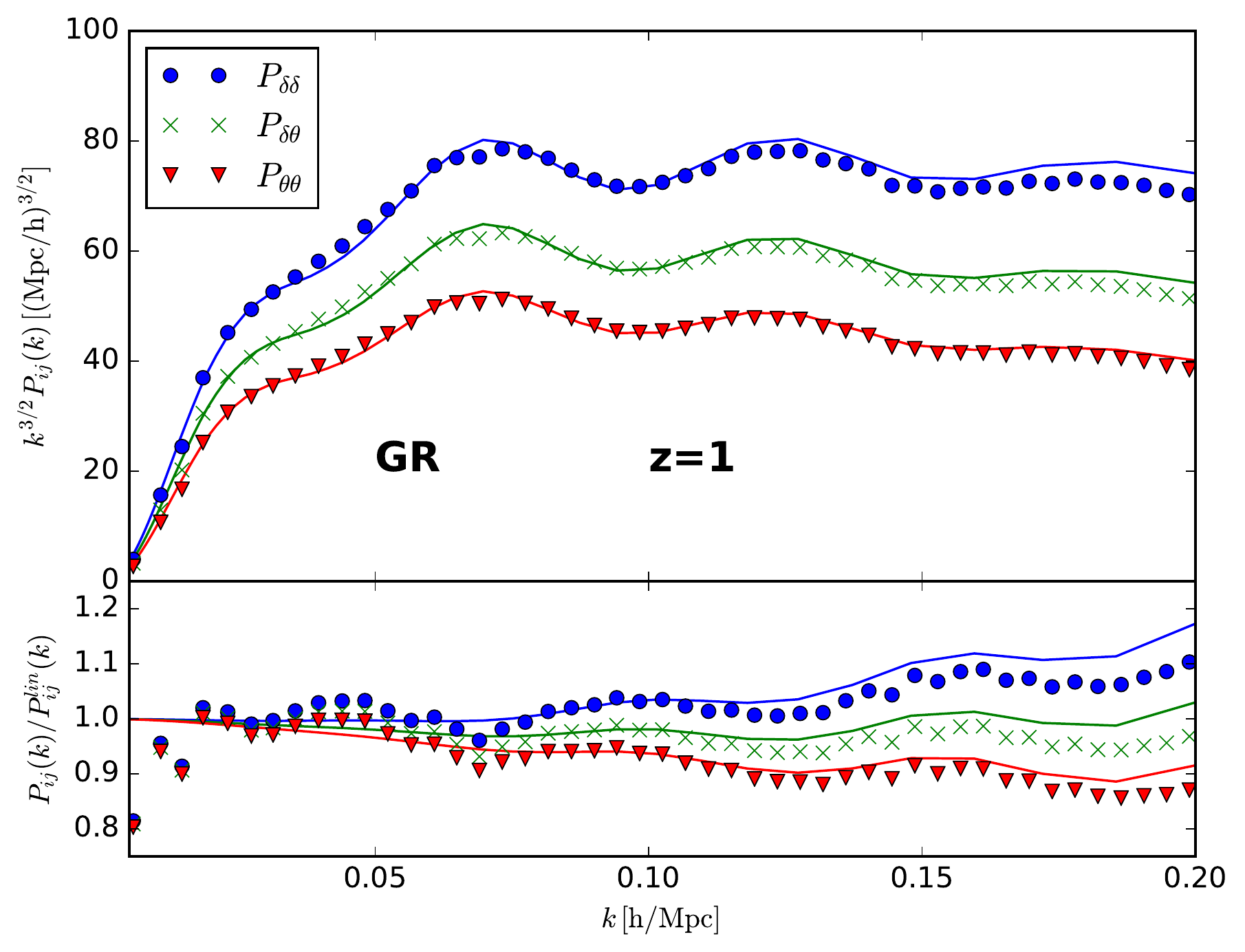}} \quad
  \subfloat[]{\includegraphics[width=8.3cm, height=8.3cm]{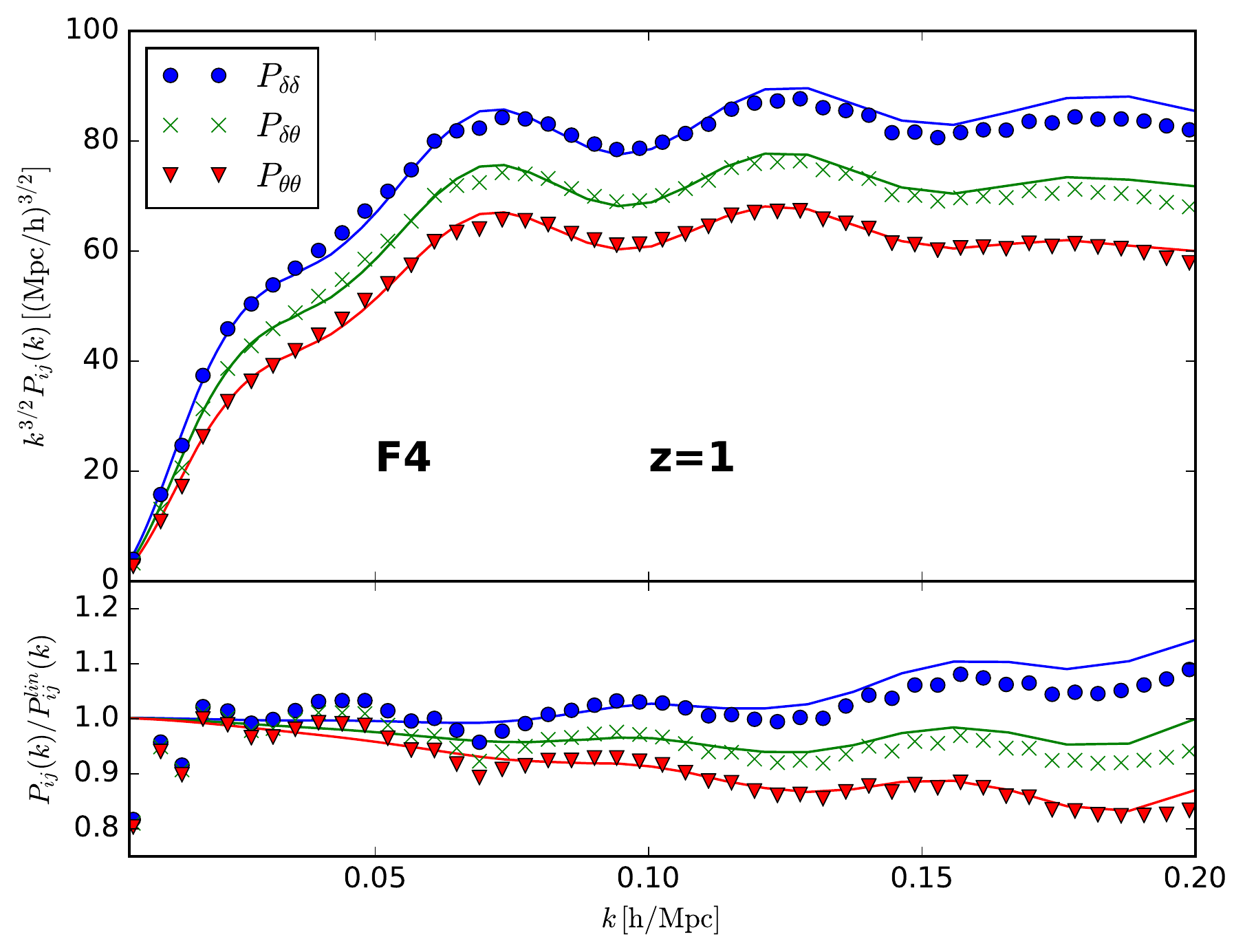}} 
  \caption[CONVERGENCE ]{Comparing our code's predictions with the N-body predictions of the auto and cross power spectra of density and velocity fields in real space at $z=1$ for GR (left) and $f(R)$ (right).  The top panels show the power spectra multiplied by $k^{3/2}$ and the bottom panels show the deviations from the linear predictions.  }
\label{frps}
\end{figure}

\begin{figure}[H]
  \captionsetup[subfigure]{labelformat=empty}
  \centering
  \subfloat[]{\includegraphics[width=8.3cm, height=8.3cm]{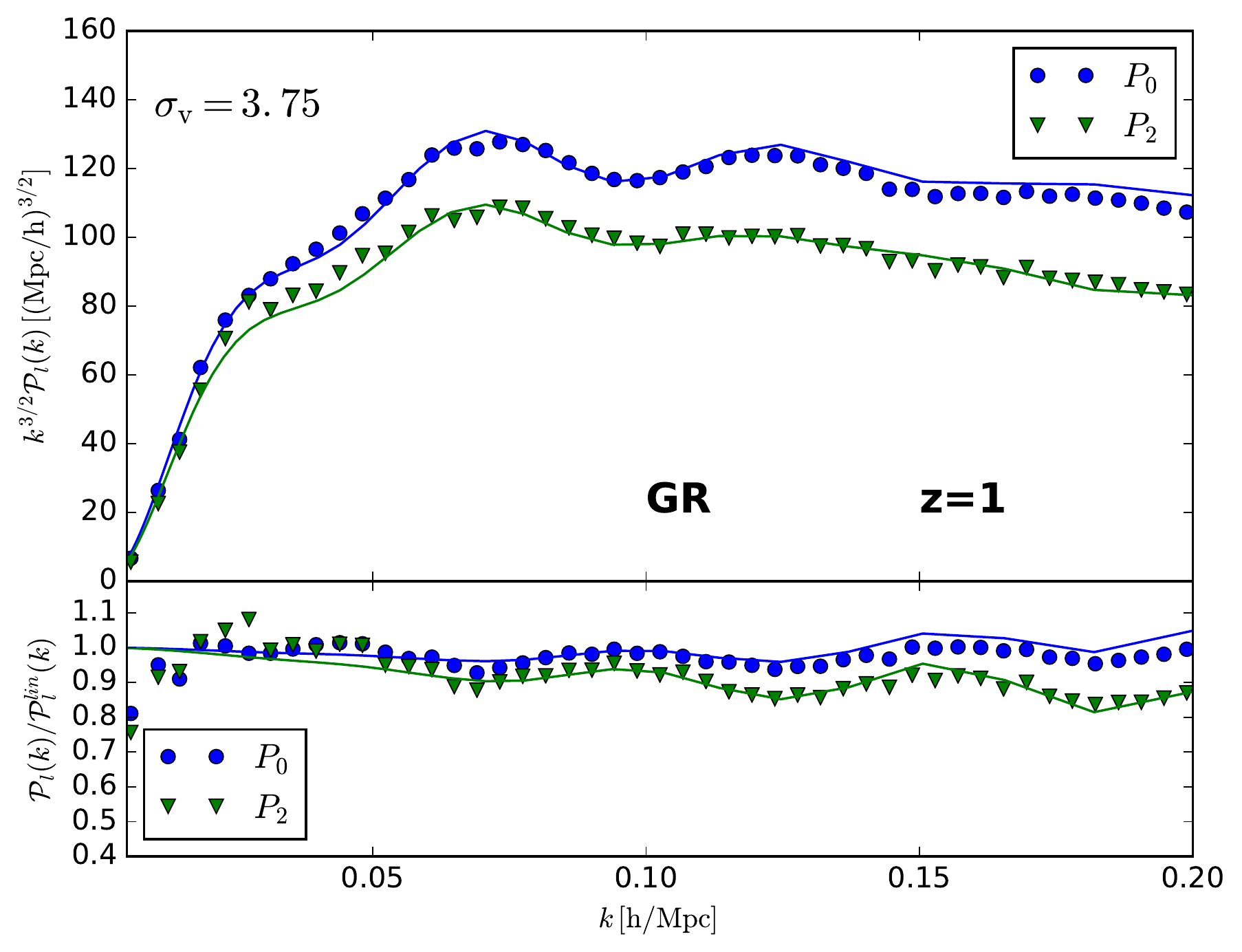}} \quad
  \subfloat[]{\includegraphics[width=8.3cm, height=8.3cm]{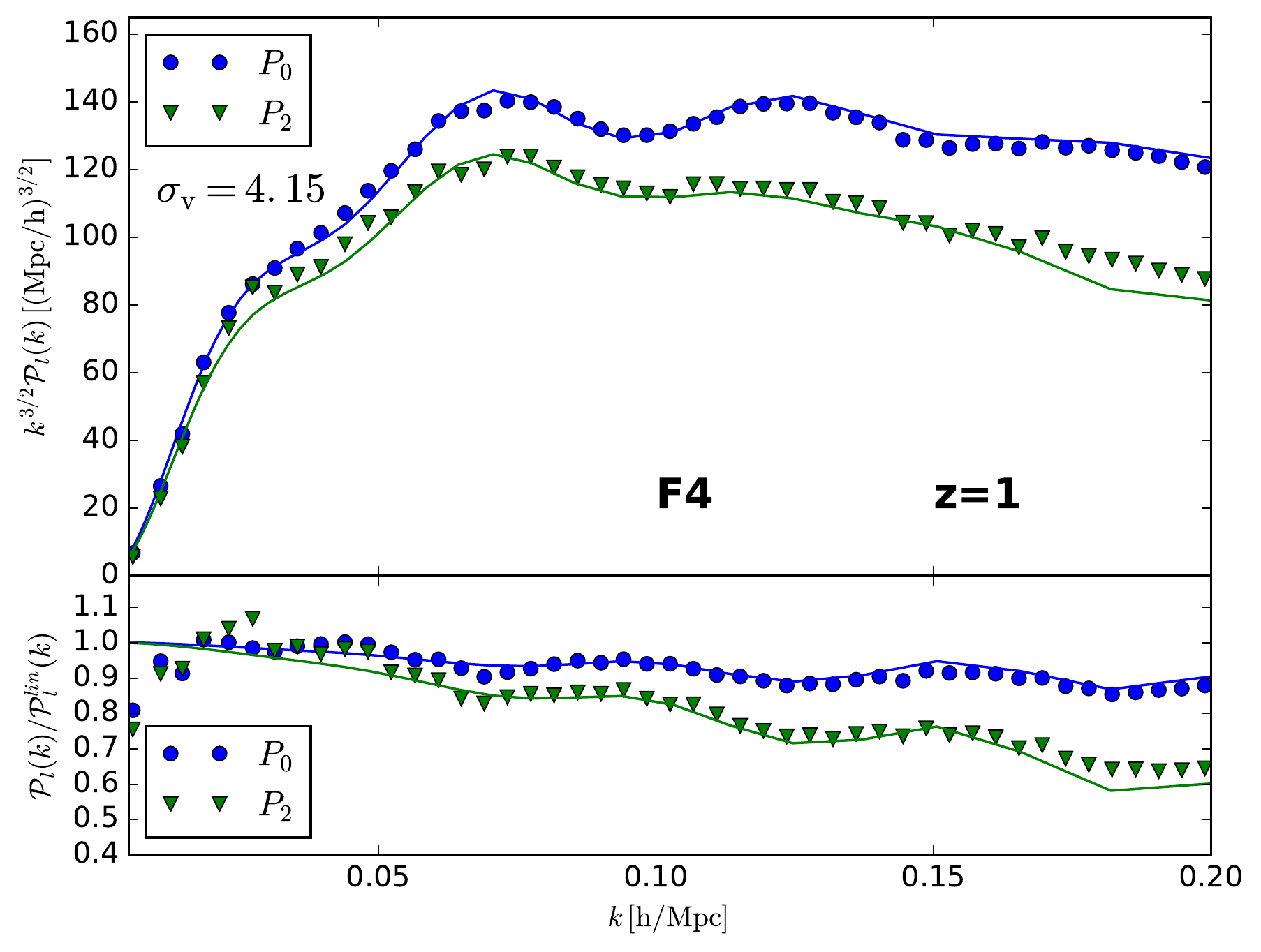}} 
  \caption[CONVERGENCE ]{Comparing our code's predictions with the N-body predictions of the TNS redshift space monopole (blue) and quadrupole (green) power spectra at $z=1$ for GR (left) and $f(R)$ (right).  The top panels show the multipoles multiplied by $k^{3/2}$ and the bottom panels show the deviations from the linear predictions. }
\label{frmulti}
\end{figure}

\begin{figure}[H]
  \captionsetup[subfigure]{labelformat=empty}
  \centering
  \subfloat[]{\includegraphics[width=10.5cm, height=8.3cm]{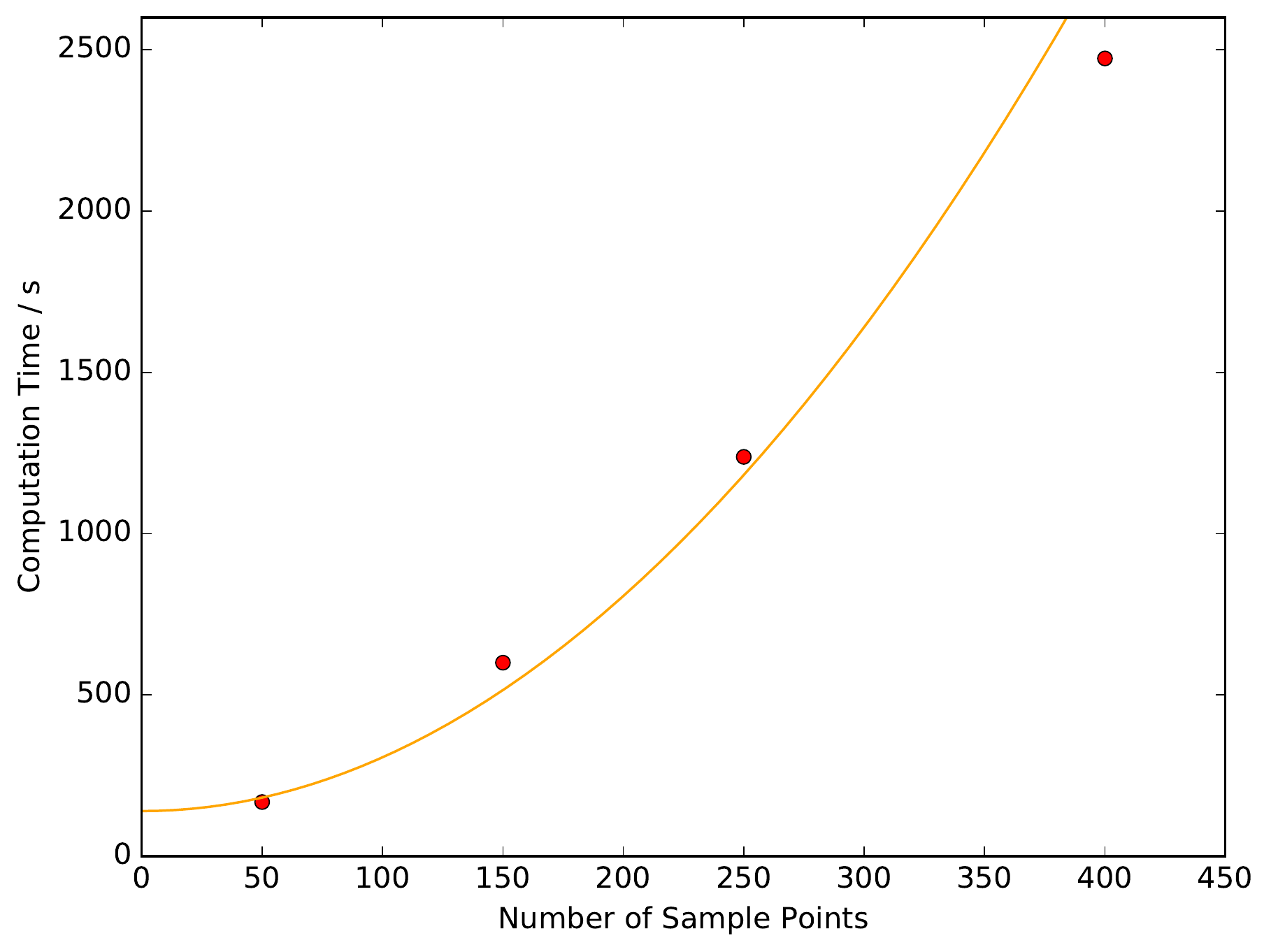}} \quad
  \caption[CONVERGENCE ]{Computation time of 30 modes of the 1-loop power spectra for multiple initialisations of the kernels. A quadratic curve is fitted to the data points.}
\label{frtime}
\end{figure}

\begin{figure}[H]
  \captionsetup[subfigure]{labelformat=empty}
  \centering
  \subfloat[]{\includegraphics[width=10.5cm, height=8.3cm]{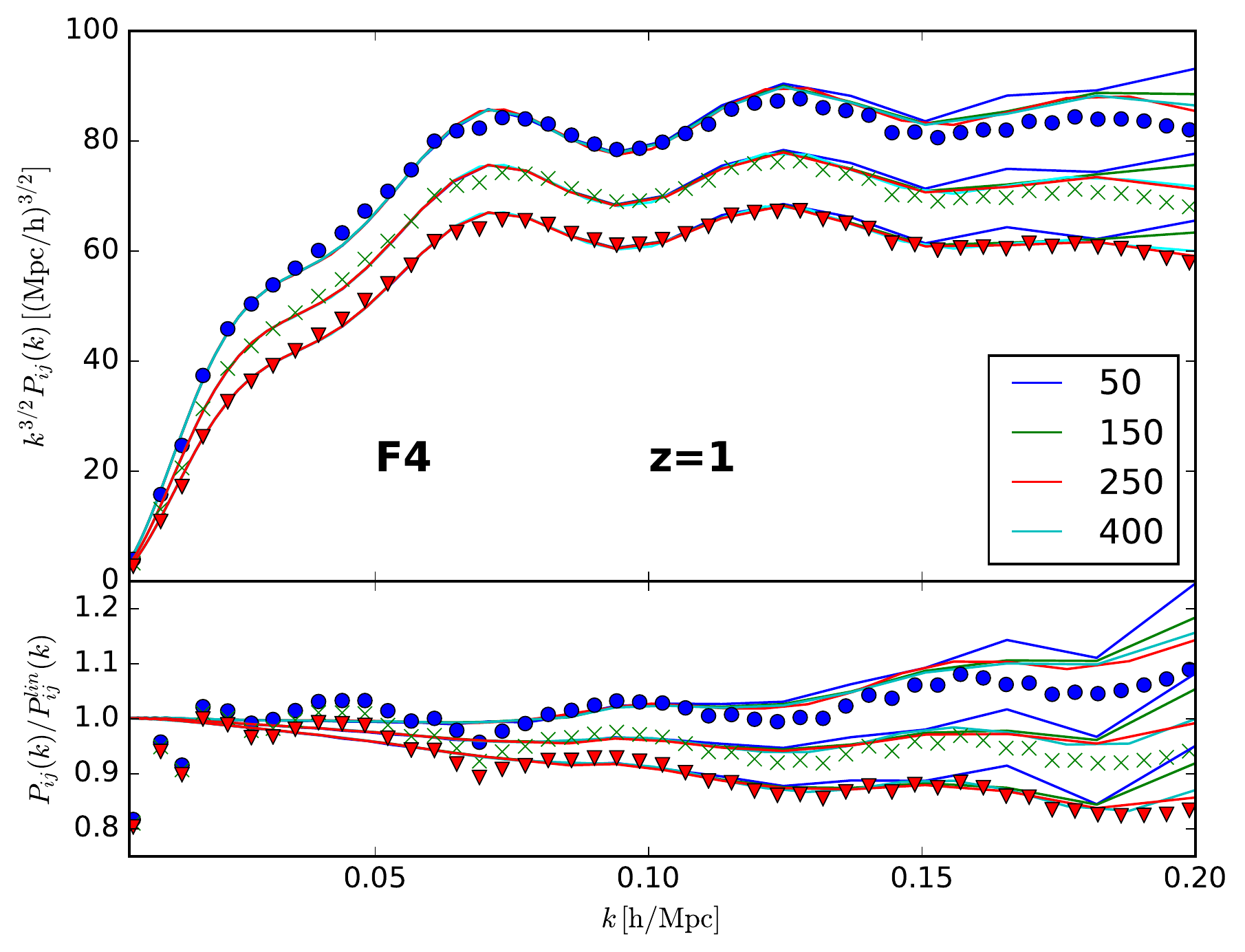}} \quad
  \caption[CONVERGENCE ]{Replica of Fig.\ref{frps} (right)  for for $n1=n2=50, 150, 250$ and $400$.}
\label{frconv}
\end{figure}

Here we attempt to reproduce some results of Ref.\cite{Taruya:2013quf}. In that work the authors compute the  SPT predictions for the 1-loop power spectra as well as the TNS redshift space monopole and quadrupole moments  for $f(R)$ and compare them with N-body results. The N-body results they use are from a subset of simulations presented in Refs.\cite{Li:2012by},\cite{Jennings:2012pt}. They use cubic boxes of size $1.5 $ Gpc/h and $1024^3$ particles, with initial conditions generated at $z=49$ using the linear matter power spectrum generated with the parameters $\Omega_m=0.24, \Omega_\Lambda = 0.76, \Omega_b=0.0481,$ h$=0.73, n_s=0.961$ and $\sigma_8=0.801$. These parameters determine the linear power spectrum we use for the SPT predictions.  
\newline
\newline
The authors consider the model of Hu and Sawicki with $n=1$ and $|f_{R0}|=10^{-4}$, henceforth called $F4$.  The results are computed at $z=1$ corresponding to $a=0.5$. 
\newline
\newline
Fig.\ref{frps} compares the real space power spectra ($P_{\delta \delta}, P_{\delta \theta}$ and $P_{\theta \theta}$) from our SPT code (computed at 1-loop order) with the N-body results for both GR and F4. We find the N-body results match the SPT calculations very well within the considered range of scales. The bottom panels of the plot also compare the non-linearity coming from the 1-loop corrections with the full non linearity of the simulations. The corrections agree with the N-body results at the percent level up to $k\sim 0.12$ h/Mpc and  the 3\% level at $k\sim 0.15$ h/Mpc.  We note here that the range of validity of SPT in $f(R)$ as dictated by eq.(\ref{validityrange}) is narrower. This is because we have stronger growth at linear order because of fifth force effects. Fig.\ref{frps} and Fig.\ref{frmulti} both exhibit the earlier breakdown of the SPT framework in this model. 
\newline
\newline
Moving to redshift space, we look at the monopole and quadrupole in the TNS model. For the FoG function we adopt a Gaussian damping function 
\begin{equation}
D_{FoG}(k\mu f \sigma_v) = exp[-(k\mu f \sigma_v)^2] 
\end{equation}
where $\sigma_v$ is a free parameter to be fit to the N-body data and $f = d \ln{F_1(k;a)}/d \ln{a}$ is the logarithmic growth factor (Note that by using the 1st order Continuity equation, we can write $f = G_1(k;a)/F_1(k;a)$). We fit $\sigma_v$ using a $\chi^2$ fit
\begin{equation}
\chi^2 = \sum_{l=0,2}\sum_i\frac{ \left[P_{l,N-body}^{(S)}(k_i)-P_{l,SPT}^{(S)}(k_i)\right]^2}{[\Delta P_{l}^{(S)}(k_i)]^2}
\end{equation}
where we sum up to $k_i=0.15$h/Mpc. We find that for F4 $\sigma_v=4.15$Mpc/h while for GR we find a lower best fit, $\sigma_v=3.75$Mpc/h. This is expected because of fifth force enhancements of velocities. Fig.\ref{frmulti} shows that the TNS model can accurately fit the data within the realm of validity of SPT for both models. We note some deviations in the quadrupole at low $k$, but lower statistics in this regime increase the error on the N-body points (not shown in the plot) keeping the agreement within the desired accuracy. 
\newline
\newline
Finally we look at the time cost of initialising the kernels for each $k$ mode in $f(R)$ gravity.  In this case the range of $r$ is smaller than in the single initialisation case and so $n2$ can be reduced with the same fineness in sampling. From Fig.\ref{frtime} and Fig.\ref{frconv} we see that increasing the sampling gives a large time cost for very little improvement in accuracy when compared to the N-body predictions. The time cost for an output of 30 values of $P_{\delta \delta}, P_{\delta \theta}$ and $P_{\theta \theta}$ at 1-loop level for $n1 = n2 = 150$ is 10 minutes on the laptop described in the previous section.  For the results in this section we used $n1=n2=200$. As previously mentioned the code is highly parallelisable and has been both MPI and Openmp enabled. 

\newpage
\newpage
\newpage
\section{Summary}
In this paper we have introduced a code which can calculate the redshift space power spectrum up to 1-loop order and its multipoles using a numerical method recently proposed in Ref.\cite{Taruya:paper}. The calculation can easily be applied to a large class of modified gravity and dark energy models through the Poisson equation source terms. These include both Vainshtein and Chameleon screened models. An example of each of these, namely the DGP and Hu- Sawicki models, were considered. The code will prove to be a very useful tool when comparing various gravitational/dark energy model's predictions against the real high precision data of ongoing and upcoming spectroscopic surveys such as the Dark Energy Spectroscopic Instrument (DESI) \footnote{\url{http://desi.lbl.gov/}}, Subaru Measurement of Imaging and Redshifts \footnote{\url{http://sumire.ipmu.jp/en/}} and the ESA/Euclid survey\footnote{\url{www.euclid-ec.org}}. All these surveys will be able to provide precise measures of the power spectrum. 
\newline
\newline
We primarily considered the TNS model of RSD because it accounts for higher order interactions. The non-linear nature allows us to properly accommodate the non-linear effects of the model which contribute a significant signature to structure growth. It also gives us a larger working range of scales in which to test the theory but still far away from the screened regime where there should be no signature of modified gravity. 
\newline
\newline
The code is written in the {\sc c++} language and is based off the cosmological perturbation theory code Copter \cite{copter}. It uses the numerical algorithm presented in Ref.\cite{Taruya:paper} to calculate the SPT kernels which is applicable to the class of gravity models we consider. Our employment of the algorithm is shown to be consistent with the exact analytical expressions in the EDS cosmology. Similar sub percent matches for  the LCDM and nDGP models may give information on  the validity of the separability ansatz that is used in the analytic solutions. The validity of this ansatz becomes more relevant on the scales considered and at the redshift considered. We have found that at smaller redshifts further, albeit still sub percent, deviations from analytic results become manifest. The biggest deviations are in the velocity velocity power spectrum, reaching about 0.5\% deviation (with $n_1 = n_2 = 800$) at $k \sim 0.2$ h/Mpc for the LCDM cosmology. These claims of validity all assume a trust in the numerical code which is strengthened by the EDS consistency check. In Ref.\cite{Fasiello:2016qpn} the authors derive an exact treatment of the density and velocity time dependence instead of using the usual EDS approximation. They find that the exact solution, in a similar-to-our LCDM cosmology, for the density density 1-loop power spectrum is larger than the EDS approximated solution at the sub 0.5\% level and is above 1\% for the velocity density cross 1-loop power spectrum at $k \sim 0.2$ h/Mpc at $z=0$. These results are consistent with Fig.\ref{convergence2}. 
\newline
\newline
 From the convergence graphs in the previous section we note that increasing the sampling numbers gives a convergence of the numerical solution giving us a handle on the error arising from finite sampling. Increased sampling comes at a computation time cost and we find good accuracy can be achieved for relatively low sampling.  We remind that this cost is easily alleviated through parallelisation of the independent $k$ modes. The cost of the 1-loop SPT calculations is presented in Fig.\ref{frtime} and Fig.\ref{time1} which show one can get high accuracy results on a laptop with fair specifications without the need for parallelisation. 
\newline
\newline
In this paper we successfully replicated results selected from Ref.\cite{Taruya:2013quf}, which looks at the Chameleon screened Hu-Sawicki model. A current work in progress is conducting the same analysis for the Vainshtein screened DGP model using the code presented here. Beyond this,  extending the range in $k$ space of the SPT calculations is also very desirable. Many extentsions to SPT already exist on the market (see Ref.\cite{Carlson:2009it}). We look to extend the range of the perturbative scheme in future work and to implement the extensions into the code.  

\section*{Acknowledgments}
\noindent We would like to thank Atsushi Taruya for useful guidance whilst writing this code and sharing his paper \cite{Taruya:paper} before its publication. We are also thankful to Elise Jennings for providing the N-body data used in Sec. IV. BB is supported by the University of Portsmouth. KK is supported by the European Research Council through 646702 (CosTesGrav).
\newpage
\appendix
\section{Einstein-de Sitter Kernels} 
Here we present the Einstein-de Sitter forms of the density and velocity divergence field kernels used as initial conditions in our code  (See Ref.\cite{Bernardeau:2001qr} for a derivation). The second order kernels are given by 
\begin{align}
F_2(\bfk_1,\bfk_2) &= \frac{5}{14} \{ \alpha(\bfk_1,\bfk_2) + \alpha(\bfk_2,\bfk_1)\} + \frac{1}{7}\beta(\bfk_1,\bfk_2) \nonumber \\
G_2(\bfk_1,\bfk_2) &= \frac{3}{14} \{ \alpha(\bfk_1,\bfk_2) + \alpha(\bfk_2,\bfk_1)\} + \frac{2}{7}\beta(\bfk_1,\bfk_2)
\nonumber 
\end{align}
where $\alpha(\bfk_1,\bfk_2)$ and $\beta(\bfk_1,\bfk_2)$ are given in eq.(\ref{alphabeta}). The 3rd order kernels are given by

\begin{align}
F_3(\bfk_1,\bfk_2,\bfk_3) = \frac{1}{3} &\left[ \frac{2}{63} \beta(\bfk_1,\bfk_{23}) \Big\{\beta(\bfk_2,\bfk_3)+\frac{3}{4}\left(\alpha(\bfk_2,\bfk_3)+\alpha(\bfk_3,\bfk_2)\right) \Big\} \right.  \nonumber \\
& + \frac{1}{18} \alpha(\bfk_1,\bfk_{23}) \Big\{\beta(\bfk_2,\bfk_3)+\frac{5}{2}\left(\alpha(\bfk_2,\bfk_3)+\alpha(\bfk_3,\bfk_2)\right)\Big\}  \nonumber \\
& \left. + \frac{1}{9} \alpha(\bfk_{23},\bfk_1) \Big\{\beta(\bfk_2,\bfk_3)+\frac{3}{4}\left(\alpha(\bfk_2,\bfk_3)+\alpha(\bfk_3,\bfk_2)\right) \Big\}  + (\mbox{cyclic perm.}) \right]
\end{align}

\begin{align}
G_3(\bfk_1,\bfk_2,\bfk_3) = \frac{1}{3}& \left[ \frac{2}{21} \beta(\bfk_1,\bfk_{23}) \Big\{\beta(\bfk_2,\bfk_3)+\frac{3}{4}\left(\alpha(\bfk_2,\bfk_3)+\alpha(\bfk_3,\bfk_2)\right)\Big\} \right.  \nonumber \\
& + \frac{1}{42} \alpha(\bfk_1,\bfk_{23}) \Big\{\beta(\bfk_2,\bfk_3)+\frac{5}{2}\left(\alpha(\bfk_2,\bfk_3)+\alpha(\bfk_3,\bfk_2)\right)\Big\}  \nonumber \\
&\left.+ \frac{1}{21} \alpha(\bfk_{23},\bfk_1) \Big\{\beta(\bfk_2,\bfk_3)+\frac{3}{4}\left(\alpha(\bfk_2,\bfk_3)+\alpha(\bfk_3,\bfk_2)\right)\Big\}  + (\mbox{cyclic perm.}) \right] 
\end{align}

\section{Modified Gravity Parameters in Horndeski's Theory}
Here we consider Horndeski's theory with a generalised potential. We start by again writing out the perturbed Friedman-Robertson-Walker metric in the Newtonian gauge.
\begin{equation}
ds^2 = - (1 + 2 \Phi) dt^2 +  a(t)^2 (1 - 2 \Psi) \delta_{ij} 
dx^i dx^j
\end{equation}
The equations of motion for the metric and scalar field perturbations are given by  (See Refs.\cite{Takushima:2015iha,Takushima:2013foa,Kimura:2011dc,DeFelice:2011hq} for $M_1 = M_2 = M_3 =0$ case) 

\begin{align}
-k^2&({\cal F}_{\cal T} \Psi(\bfk;a) -{\cal G}_{\cal T} \Phi(\bfk;a) - A_1 Q(\bfk;a)) \nonumber \\ 
= & \frac{B_1}{2a^2H^2}\Gamma[\bfk,Q,Q;a] + \frac{ B_3}{a^2 H^2} \Gamma[\bfk,Q,\Phi;a] \label{eomhorn1} \\ \nonumber \\ 
-k^2&({\cal G}_{\cal T} \Psi(\bfk;a) + A_2 Q(\bfk;a)) -\frac{a^2}{2}\rho_m \delta(\bfk;a) \nonumber \\
 =& -\frac{B_2}{2a^2H^2}\Gamma[\bfk,Q,Q;a] - \frac{ B_3}{a^2 H^2} \Gamma[\bfk,Q,\Psi;a] -\frac{C_1}{3a^4H^4}\Xi_1[\bfk,Q,Q,Q;a] \label{eomhorn2}   \\ \nonumber \\ 
-k^2&(A_0 Q(\bfk;a)-A_1 \Psi(\bfk;a) - A_2 \Phi(\bfk;a)) \nonumber \\
  =& -\frac{B_0}{a^2H^2}\Gamma[\bfk,Q,Q;a] + \frac{ B_1}{a^2 H^2} \Gamma[\bfk,Q,\Psi;a] +  \frac{ B_2}{a^2 H^2} \Gamma[\bfk,Q,\Phi;a] + \frac{ B_3}{a^2 H^2} \Gamma[\bfk,\Psi,\Phi;a]  \nonumber \\
& +\frac{C_0}{3a^4H^4}\Xi_1[\bfk,Q,Q,Q;a] +\frac{C_1}{3a^4H^4}\Xi_1[\bfk,Q,Q,\Phi;a] \nonumber \\
& + M_1Q(\bfk;a) + \frac{M_2}{a^2H^2} Q(\bfk;a)^2 + \frac{M_3}{a^4H^4}Q(\bfk;a)^3
\label{eomhorn3}
\end{align}
where we have parametrized the scalar field perturbation as $Q=\delta \phi/ (a\phi')$, prime denoting scale factor derivative, and we have defined
\begin{align}
\Gamma[\bfk,Z_1,Z_2;a] =& \frac{1}{(2\pi)^3}\int d^3\bfk_1 d^3 \bfk_2 \delta_D(\bfk-\bfk_{12})\lambda(\bfk_1,\bfk_2)  Z_1(\bfk_1;a) Z_2(\bfk_2;a) \\ 
\Xi_1[\bfk,Z_1,Z_2,Z_3;a] = & \frac{1}{(2\pi)^6}\int d^3\bfk_1 d^3 \bfk_2 d^3\bfk_3\delta_D(\bfk-\bfk_{123}) \nonumber \\
& \times \left[-k_1^2 k_2^2 k_3^2 + 3k_1^2(\bfk_2\cdot\bfk_3)^2-2(\bfk_1\cdot \bfk_2)(\bfk_2 \cdot \bfk_3) (\bfk_3 \cdot \bfk_1) \right] Z_1(\bfk_1;a) Z_2(\bfk_2;a)Z_3(\bfk_3;a) \\
\Xi_2[\bfk,Z_1,Z_2,Z_3;a] = & \frac{1}{(2\pi)^6}\int d^3\bfk_1 d^3 \bfk_2 d^3\bfk_3\delta_D(\bfk-\bfk_{123}) \left[ -k_1^2 k_2^2 k_3^2 + k_3^2(\bfk_1\cdot\bfk_2)^2 \right. \nonumber \\
&  \left. +2k_1^2(\bfk_2 \cdot \bfk_3)^2 -2(\bfk_1\cdot \bfk_2)(\bfk_2 \cdot \bfk_3) (\bfk_3 \cdot \bfk_1) \right] Z_1(\bfk_1;a) Z_2(\bfk_2;a)Z_3(\bfk_3;a) 
\end{align}
where $\bfk_{ijk}=\bfk_i+\bfk_j+\bfk_k$ and $ \lambda(\bfk_1, \bfk_2) =k_1^2 k_2^2 - (\bfk_1 \cdot \bfk_2)^2 $. Scalar field potential terms, $M_1(a),M_2(a)$ and $M_3(a)$, have been introduced into the scalar field perturbation's equation of motion so that we can accommodate chameleon models such as $f(R)$ gravity.  The functions $A_0,A_1,A_2,B_0,B_1,B_2,B_3,C_0,C_1$ depend only on the scale factor. In Ref.\cite{Takushima:2015iha}, these equations together with the continuity and Euler equations are solved perturbatively up to third order using the separability ansatz. Here instead we find the SPT kernels by solving eq.\ref{eq:dynamicsystem} numerically. For this purpose,  we follow a perturbative approach where we assume $\delta \ll 1$ and then expand our field perturbations, $\Psi,\Phi$ and $Q$, in terms of increasing orders of $\delta$ in order to find $\mu(k;a)$, $\gamma_2(\bfk, \bfk_1,\bfk_2;a)$ and$\gamma_3(\bfk, \bfk_1,\bfk_2,\bfk_3;a)$ of eq.\ref{eq:Perturb3} and eq.\ref{eq:poisson1}.  At first order the solutions for the metric perturbations and the scalar field under the assumption of quasi-static approximations are given by
\begin{align}
\Phi_1 & = -\frac{{\cal R}}{{\cal Z}} \frac{a^2}{k^2} \rho_m \delta 
\equiv  \frac{\rho_m \delta}{\Upsilon_{\Phi}(k)} \\
\Psi_1 &=-\frac{{\cal S}}{{\cal Z}} \frac{a^2}{k^2} \rho_m \delta 
 \equiv  \frac{\rho_m \delta}{\Upsilon_{\Psi}(k)}  \\
Q_1 &= -\frac{{\cal T}}{{\cal Z}}
 \frac{a^2}{k^2} \rho_m \delta 
\equiv  \frac{\rho_m \delta}{\Upsilon_{Q}(k)}
\end{align}
where we have introduced the quantities $\Upsilon_{I}(k)$, $I=\Phi, \Psi$ and $Q$, defined using the expressions
\begin{align}
{\cal R}= \tilde{A}_0 {\cal F}_{\cal T} - A_1^2, \quad 
{\cal S}= \tilde{A}_0 {\cal G}_{\cal T} + A_1 A_2, \quad 
{\cal T}= A_1 {\cal G}_{\cal T} +  A_2 {\cal F}_{\cal T}, \quad 
{\cal Z}= 2 (A_2^2 {\cal F}_{\cal T} + 2 A_1 A_2 {\cal G}_{\cal T} + \tilde{A}_0 {\cal G}_{\cal T}^2)
\end{align}
and
\begin{equation}
\tilde{A}_0 = A_0 + \frac{M_1}{k^2}. 
\end{equation}

At the second order, the solutions are given in the form 
\begin{align}
\Phi &= 
\int\frac{d^3\bfk_1d^3\bfk_2}{(2\pi)^3}\,
\delta_{\rm D}(\bfk-\bfk_{12}) \Gamma_{2 \Phi} (\bfk, \bfk_1, \bfk_2, k)
\delta(\bfk_1)\,\delta(\bfk_2) \\
\Psi &=
\int\frac{d^3\bfk_1d^3\bfk_2}{(2\pi)^3}\,
\delta_{\rm D}(\bfk-\bfk_{12}) \Gamma_{2 \Psi} (\bfk, \bfk_1, \bfk_2, k)
\delta(\bfk_1)\,\delta(\bfk_2) \\
Q &=
\int\frac{d^3\bfk_1d^3\bfk_2}{(2\pi)^3}\,
\delta_{\rm D}(\bfk-\bfk_{12}) \Gamma_{2 Q} (\bfk, \bfk_1, \bfk_2, k)
\delta(\bfk_1)\,\delta(\bfk_2) 
\end{align}
where
\begin{align}
\Gamma_{2 I}(\bfk, \bfk_1, \bfk_2) 
&=
\frac{\rho_m^2}{a^4H^2} \Big[ \left(
\frac{W_{1 I}(k)}{\Upsilon_{\Psi}(k_1) \Upsilon_{Q}(k_2)} 
+ \frac{W_{2 I}(k)}{\Upsilon_{\Phi}(k_1) \Upsilon_{Q}(k_2)}  
+ \frac{W_{3 I}(k)}{\Upsilon_{\Psi}(k_1) \Upsilon_{\Phi}(k_2)}  
+\frac{W_{4 I}(k)}{\Upsilon_{Q}(k_1) \Upsilon_{Q}(k_2)} 
\right) 
\lambda(\bfk_1, \bfk_2) \nonumber \\
+& \frac{W_{5 I}(k)}{\Upsilon_{Q}(k_1) \Upsilon_{Q}(k_2)} M_2
\Big],
\end{align}
where $k=k_{12}$ and 
\begin{align}
W_{1 \Phi}(k) &= \frac{2 a^2}{k^2} 
\frac{B_1 {\cal T} + B_3 {\cal R}}{{\cal Z}}, \quad
W_{2 \Phi}(k) = \frac{2 a^2}{k^2}
\frac{B_2 {\cal T} + B_3 {\cal S}}{{\cal Z}}, \quad
W_{3 \Phi}(k) = \frac{2 a^2 }{k^2}
\frac{B_3 {\cal T}}{{\cal Z}}
 \\
W_{4 \Phi}(k) &=\frac{a^2}{k^2}
\frac{-2 B_0 {\cal T}+ B_1 {\cal S} + B_2 {\cal R} }{{\cal Z}},\quad
W_{5 \Phi}(k) =\frac{2 a^2}{k^2}
\frac{{\cal T}}{{\cal Z}} \\
W_{1 \Psi}(k) &= \frac{2 a^2}{k^2} 
\frac{ A_2 B_1 {\cal G}_{\cal T} +  B_3 {\cal S}}{{\cal Z}}, \quad
W_{2 \Psi}(k) = \frac{2 a^2}{k^2}
\frac{A_2 B_2 {\cal G}_{\cal T} - A_2^2 B_3}{{\cal Z}}, \quad
W_{3 \Psi}(k) = \frac{2 a^2}{k^2}
\frac{A_2 B_3 {\cal G}_{\cal T}}{{\cal Z}}
\\
W_{4 \Psi}(k) &=\frac{a^2}{ k^2}
\frac{ - 2 A_2 B_0 {\cal G}_{\cal T} - A_2^2 B_1 + B_2 {\cal S} }{{\cal Z}},\quad
W_{5 \Psi}(k) =\frac{2 a^2}{k^2}
\frac{ A_2 {\cal G}_{\cal T}}{{\cal Z}}, \\
W_{1 Q}(k) &= \frac{2 a^2}{k^2} 
\frac{ -B_1 {\cal G}_{\cal T} + B_3 {\cal T}}{{\cal Z}}, \quad
W_{2 Q}(k) = \frac{2 a^2}{k^2}
\frac{-B_2 {\cal G}_{\cal T}^2 + A_2 B_3 {\cal G}_{\cal T}}{{\cal Z}}, \quad
W_{3 Q}(k) = -\frac{2 a^2}{k^2}
\frac{B_3 {\cal G}_{\cal T}^2}{{\cal Z}}
\\
W_{4 Q}(k) &=\frac{a^2 }{ k^2}
\frac{ 2 B_0 {\cal G}_{\cal T}^2 +  A_2 B_1 {\cal G}_{\cal T} +B_2 {\cal T}}{{\cal Z}},\quad
W_{5 Q}(k) =-\frac{2 a^2 }{k^2}
\frac{{\cal G}_{\cal T}^2}{{\cal Z}}, 
\end{align} 
At the third order 
\begin{equation}
\Phi=\int\frac{d^3\bfk_1d^3\bfk_2d^3\bfk_3}{(2\pi)^6}
\delta_{\rm D}(\bfk-\bfk_{123})
\Gamma_{3 \Phi}(\bfk, \bfk_1, \bfk_2, \bfk_3)
\delta(\bfk_1)\,\delta(\bfk_2)\,\delta(\bfk_3)
\end{equation}
where
\begin{align}
\Gamma_{3 \Phi}(\bfk, \bfk_1, \bfk_2, \bfk_3)
&= \frac{ \rho_m^3}{3a^6H^4 } 
\left[    \frac{Z_0(k)}{\Pi_Q(k_1) \Upsilon_{Q}(k_2) \Upsilon_{\Phi}(k_3)} \xi (\bfk_1, \bfk_2, \bfk_3) + (\mbox{cyclic perm.}) \right.
 \nonumber  \\
&+\left.
\frac{3Z_1(k)}{\Pi_Q(k_1) \Pi_Q(k_2) \Pi_Q(k_3)}
\xi (\bfk_1, \bfk_2, \bfk_3)
+\frac{3Z_2(k)}{\Pi_Q(k_1) \Pi_Q(k_2) \Pi_Q(k_3)} M_3 \right] \nonumber \\
&  + \frac{\rho_m}{3a^4H^2} \left[ Z_3(k) \frac{\lambda(\bfk_1, \bfk_{23}, k) \Gamma_{2 \Psi} (\bfk, \bfk_2, \bfk_3, k_{23})}{\Upsilon_{\Phi}(k_1)}   + 
Z_4(k) \frac{\lambda(\bfk_1, \bfk_{23}, k) \Gamma_{2 \Psi} (\bfk, \bfk_2, \bfk_3, k_{23})}{\Upsilon_{Q}(k_1)} \right.  \nonumber \\
&\left. +
Z_5(k) \frac{\lambda(\bfk_1, \bfk_{23}, k) \Gamma_{2 \Phi} (\bfk, \bfk_2, \bfk_3, k_{23})}{\Upsilon_{\Psi}(k_1)} 
+ 
Z_6(k) \frac{\lambda(\bfk_1, \bfk_{23}, k) \Gamma_{2 \Phi} (\bfk, \bfk_2, \bfk_3, k_{23})}{\Upsilon_{Q}(k_1)} \right. \nonumber \\
& \left.+ 
Z_7(k) \frac{\lambda(\bfk_1, \bfk_{23}, k) \Gamma_{2 Q} (\bfk, \bfk_2, \bfk_3, k_{23})}{\Upsilon_{\Phi}(k_1)} 
+
Z_8(k) \frac{\lambda(\bfk_1, \bfk_{23}, k) \Gamma_{2 Q} (\bfk, \bfk_2, \bfk_3, k_{23})}{\Upsilon_{\Psi}(k_1)} \right. \nonumber  \\
&\left. +
Z_9(k) \frac{\lambda(\bfk_1, \bfk_{23}, k) \Gamma_{2 Q} (\bfk, \bfk_2, \bfk_3, k_{23})}{\Upsilon_{Q}(k_1)} 
+
Z_{10}(k) \frac{M_2 \Gamma_{2 Q} (\bfk, \bfk_2, \bfk_3, k_{23})}{\Upsilon_{Q}(k_1)} + (\mbox{cyclic perm.}) \right] 
\end{align}
where
\begin{align}
Z_0(k) &= \frac{2 a^2}{k^2} \frac{C_1 {\cal T}}{{\cal Z}}, \quad  
Z_1(k) =  \frac{2a^2 }{3k^2} \frac{3 C_0 {\cal T}
+ C_1 {\cal R} }{ {\cal Z}}, \quad
Z_2(k) = \frac{a^2 }{k^2} \frac{2 {\cal T}}{{\cal Z}}, \quad 
Z_3(k) = \frac{a^2 }{k^2} \frac{2 B_3 {\cal T}}{{\cal Z}}, \\
Z_4(k) &= \frac{2 a^2 }{k^2} 
\frac{B_1 {\cal T}+ B_3 {\cal R}}{{\cal Z}}, \quad 
Z_5(k) = \frac{2 a^2 }{k^2} 
\frac{B_3 {\cal T}}{{\cal Z}}, \quad 
Z_6(k) = \frac{2 a^2 }{k^2} 
\frac{B_2 {\cal T}+ B_3 {\cal S} }{{\cal Z}}, \\
Z_7(k) &= \frac{2 a^2 }{k^2} 
\frac{B_2 {\cal T}+B_3 {\cal S}}{{\cal Z}}, \quad 
Z_8(k) = \frac{2 a^2 }{k^2} 
\frac{B_1 {\cal T} + B_3 {\cal R}}{{\cal Z}}, \quad 
Z_9(k) = \frac{2 a^2 }{k^2} 
\frac{-2 B_0 {\cal T} +B_1 {\cal S}+B_2 {\cal R}}{{\cal Z}},\\
Z_{10}(k) & = \frac{4 a^2 }{k^2} 
\frac{{\cal T}}{{\cal Z}},
\end{align}
and 
\begin{align}
\xi(\bfk_1, \bfk_2, \bfk_3)
 =   
k_1^2 k_2^2 k_3^2- k_1^2 (\bfk_2 \cdot \bfk_3)^2 + k_2^2 (\bfk_3 \cdot \bfk_1)^2 + k_3^2 (\bfk_1 \cdot \bfk_2)^2 
	-2 (\bfk_1 \cdot \bfk_2) (\bfk_2 \cdot \bfk_3)(\bfk_3 \cdot \bfk_1)
\end{align}
For a given theory, we need only compare the field perturbation equations of motion (eq.(\ref{eomhorn1}), eq.(\ref{eomhorn2}) and eq.(\ref{eomhorn3})) to determine an expression for $\Phi$ in orders of $\delta$. The Euler equation modification parameters given in the first section, $\mu(k;a), \gamma_2(\bfk, \bfk_1,\bfk_2;a)$ and $\gamma_3(\bfk, \bfk_1,\bfk_2,\bfk_3;a)$ can then be related to $\Upsilon_\Phi(k;a),\Gamma_{2\Phi}(\bfk, \bfk_1,\bfk_2;a)$ and $\Gamma_{3\Phi}(\bfk, \bfk_1,\bfk_2,\bfk_3;a)$ through order by order comparisons with eq.(\ref{eq:poisson1}). We find the relations are 

\begin{align}
\mu(k;a) & = -\frac{k^2}{a^2} \frac{2}{\kappa^2 \Upsilon_\Phi(k;a)} \\
\gamma_2(\bfk, \bfk_1,\bfk_2;a) &= -\frac{k^2}{a^2H^2} \Gamma_{2\Phi} (\bfk, \bfk_1,\bfk_2;a) \\ 
\gamma_3(\bfk, \bfk_1,\bfk_2,\bfk_3;a) &=  -\frac{k^2}{a^2H^2} \Gamma_{3\Phi} (\bfk, \bfk_1,\bfk_2,\bfk_3;a) 
\end{align}
The functions on the right hand sides are expressed in terms of $A_0,A_1,A_2,B_0,B_1,B_2,B_3,C_0$ and $C_1$  which are in turn relatable to the Horndeski Lagrangian functions (See Ref.\cite{Takushima:2015iha} for example). 
\renewcommand{\bibname}{References}
\bibliography{mybib}{}
\end{document}